\documentclass[nofootinbib,twocolumn,secnumarabic,amssymb, nobibnotes, aps, prd]{revtex4-2}

\usepackage{hyperref}       
\usepackage{amsmath}
\usepackage{adjustbox}
\usepackage{orcidlink} 
\usepackage{graphicx}
\usepackage{booktabs}      
\usepackage{amsfonts}      
\usepackage{nicefrac}       
\usepackage{microtype}
\usepackage{cleveref}
\usepackage{xcolor}
\hypersetup{
    colorlinks,
    linkcolor={red!100!black},
    citecolor={green!50!black},
    urlcolor={blue!50!black}
}
\usepackage{multirow}
\usepackage{makecell}
\usepackage{array}
\usepackage{lipsum}

\newcommand{\vect}[1]{\boldsymbol{#1}}

\begin{document}

\title{On the Calculation of Pressure Derivatives in Mean-Field Thermal Field Theories}

\author{Hosein Gholami \orcidlink{0009-0003-3194-926X}}
\email{mohammadhossein.gholami@tu-darmstadt.de}
\affiliation{Technische Universität Darmstadt, Fachbereich Physik, Institut für Kernphysik,
Theoriezentrum, Schlossgartenstr.~2 D-64289 Darmstadt, Germany
}

\begin{abstract}
Accurate determination of higher-order pressure derivatives with respect to temperature \(T\) and chemical potential \(\mu\) is essential for analyzing critical phenomena, transport properties, and phase transitions in strongly interacting matter. However, standard numerical differentiation methods often suffer from large numerical instabilities, especially in more complex mean-field thermal field theories. In this work, we present an approach that systematically derives symbolic expressions for these higher-order derivatives, bypassing the numerical instabilities commonly encountered in conventional methods. Our formalism is based on a Jacobian technique, which ensures that the dependence of internal mean-field parameters is fully incorporated into the final symbolic expressions. We illustrate the effectiveness of this method using the two-flavor Nambu--Jona-Lasinio model as an example and show that it is particularly advantageous near phase transitions and at low temperatures, where numerical differentiation becomes highly sensitive.
\end{abstract}

\maketitle

\section{Introduction}

In the study of strongly interacting matter, accurately determining thermodynamic properties is essential. Quantities such as the speed of sound, heat capacities, and particle number susceptibilities depend on higher-order derivatives of the thermodynamic potential with respect to temperature $T$ and chemical potential $\mu$. These higher-order derivatives are helpful for understanding the behavior of matter in regimes explored by heavy-ion collisions, neutron star interiors, and the QCD phase diagram.

Traditionally, higher-order derivatives of the pressure or the thermodynamic potential are computed via finite-difference schemes. While conceptually straightforward, these methods are prone to numerical uncertainties. At small resolutions, finite-difference formulas suffer from rounding errors and ill-conditioning, which, when combined with noise in discrete data, can significantly inflate errors, particularly for higher-order derivatives (see, e.g., Ref.~\cite{10.5555/148286} for a general overview or Refs.~\cite{Grossi:2019urj,Koenigstein:2021syz} for finite-difference errors in functional renormalization group (FRG) calculations). This limitation underscores the need for alternative methods that avoid these pitfalls.
Ref.~\cite{Mroczek:2024sfp} highlighted these challenges in their study on finite-temperature expansions of the dense-matter equation of state, where numerical instabilities near critical points posed significant difficulties. Similarly, Ref.~\cite{Cruz-Camacho:2024odu} investigated phase stability and multidimensional phase boundaries within the Chiral Mean-Field model. Although numerical noise was not explicitly discussed, the inherent complexity of such calculations makes them particularly susceptible to these issues.
Additionally, many studies performing calculations of the speed of sound squared and heat capacity rely on numerical differentiation, which is similarly prone to these limitations. In Ref.~\cite{gholami2024astrophysicalconstraintscolorsuperconductingphases}, we encountered numerical instabilities in the calculation of the speed of sound squared. The issue arose due to the complexity of the model, where small fluctuations in the pressure and energy density led to significant variations in the calculated value of the speed of sound. To address this, we had to ensure extra precision in the computation of pressure and energy density. The method proposed here provides a robust framework to overcome these challenges and can be complementary to these works and similar works.

In the context of strongly interacting matter, mean-field 
theories approximate many-body interactions by assuming each constituent 
experiences an average background field generated by all others. Solving the 
corresponding self-consistent gap equations then provides 
the set of internal parameters (e.g., mass gaps, diquark condensates). Consequently, an 
\(n\)-th order derivative typically requires evaluating the pressure at 
\(n+1\) closely spaced points, and each evaluation often demands a full 
mean-field minimization---i.e., numerically solving the gap equations at each 
grid point. Such multiple evaluations amplify round-off errors, increase 
computational cost, and risk large numerical fluctuations. Furthermore, near 
phase transitions, where the thermodynamic potential changes rapidly, these 
finite difference calculations become particularly delicate. Data limitations, 
storage constraints, and steep gradients can lead to substantial deviations 
from true values.

It is also common practice to attempt reducing numerical noise by fitting the computed pressure data to a spline or other smooth interpolating function, then taking analytic derivatives of this fit. However, if the fit is not carefully chosen or if the interpolation order is large, it can introduce oscillations or artifacts. These inaccuracies propagate to the derived higher-order derivatives, compromising their reliability. Thus, while fitting and splining can sometimes mitigate noise, they neither eliminate the fundamental difficulties of numerical differentiation nor guarantee physically meaningful results.

To address these issues, we propose an approach that uses a Jacobian-based 
formalism to derive symbolic expressions for higher-order pressure derivatives. 
This method does not rely on multiple closely spaced pressure evaluations and 
thus avoids the accumulation of round-off and subtraction errors. By 
constructing these analytic formulae, only a single evaluation at the 
physical solution---namely, the mean-field configuration that 
minimizes the effective potential---is required. This can potentially reduce computational cost, 
removes the need for fitting or spline-based smoothing methods, and ensures 
more stable results.

As an illustrative example, we consider a standard mean-field grand canonical 
finite-temperature field theory. In this setting, the thermodynamic variables
---thermodynamic (Landau) potential \(\Omega\), energy density \(\epsilon\), 
entropy density \(s\), and number density \(n\)---depend on the external 
parameters \(\mu\) and \(T\). The pressure is defined by \(P = -\Omega\). Hence, we have
\begin{eqnarray}
\Omega(T,\mu) &=& \epsilon - T\,s - \mu\,n\quad, \label{eq:dOmega}\\
P(T,\mu) &=& -\Omega(T,\mu). 
\end{eqnarray} 
We can expand the pressure in a Taylor series around any reference point 
\((T_0,\mu_0)\). To facilitate this, we define the \emph{pressure expansion 
coefficients} as
\begin{equation} c_{m,n}(T,\mu) = \frac{\partial^{m+n} P}{\partial T^m \partial \mu^n}, \label{eq:pressure_expansion_coefficients} \end{equation}
where \(m,n \ge 0\). The order in which the derivatives are taken does not 
matter. 
Using these coefficients, we can systematically study the thermodynamic properties of strongly interacting matter, particularly in the context of the QCD phase diagram. For instance, these coefficients have been employed in Ref.~\cite{Karsch:2010hm} to analyze the convergence of Taylor expansions near phase transitions in effective models and lattice QCD, and in Ref.~\cite{Schaefer:2011ex} to investigate the scaling behavior of higher-order moments of thermodynamic susceptibilities and their connection to the QCD critical point. More recently, Ref.~\cite{Mroczek:2024sfp} applied a similar expansion framework to construct a finite-temperature equation of state for dense matter.

For instance, the first-order derivatives in Eq.~\eqref{eq:pressure_expansion_coefficients}
\begin{align}
c_{0,1} &= \frac{\partial P}{\partial \mu}\bigg|_T = n, \label{eq:c01}\qquad c_{1,0} = \frac{\partial P}{\partial T}\bigg|_\mu = s,
\end{align}
directly yield the number density and entropy density, respectively. Our primary 
focus, however, lies in higher-order coefficients (e.g., \(c_{0,2}, c_{1,1}, 
c_{2,0}\)), which tend to be more sensitive to the numerical issues mentioned 
above.

To illustrate the advantages of our method, we apply it to the two-flavor 
Nambu--Jona-Lasinio (NJL) model and its extension with diquark pairing. These 
models serve as instructive testing grounds because they involve nontrivial 
internal parameter determinations---given that the solutions to the gap 
equations are not available in closed form---and also allow investigations of 
first-order, second-order, and crossover phase transitions within the same 
theoretical framework.

The structure of this paper is as follows. In Sec.~\ref{sec:2}, we introduce the general mean-field framework and discuss the difficulties in the symbolic calculation of higher derivatives. In Sec.~\ref{sec:3}, we present the Jacobian-based formalism that yields symbolic expressions for higher-order derivatives. In Sec.~\ref{sec:examp}, we apply our method to the two-flavor NJL model and its diquark-extended version, showing how the symbolic approach resolves limitations of numerical differentiation. We conclude in Sec.~\ref{sec:conclusions} by summarizing our results and outlining potential future applications of the method.

\section{Mean-Field Thermal Field Theory}\label{sec:2}

Our goal is to establish a procedure for calculating symbolic higher-order pressure derivatives within a mean-field thermal framework. It is crucial to understand the distinction between the \emph{effective potential}, which defines the model and depends on internal parameters, and the \emph{thermodynamic potential}, which is a physically observable quantity expressed solely in terms of the external (free) variables --- e.g., \(T\) and \(\mu\) in grand canonical ensemble. Our method leverages this relationship, revealing that derivatives of the physical pressure can be understood as derivatives of the effective potential taken under suitable constraints.

In a general grand canonical thermal field theory, the \textit{effective} potential is defined as
\begin{equation}\label{eq:effpot}
\Omega^\text{eff}(T,\mu,\vect{\chi}) \equiv \frac{T}{V_3}\,\Gamma(T,\mu,\vect{\chi}),
\end{equation}
where \(V_3\) is the three-dimensional spatial volume, \(\Gamma\) is the real homogeneous part of the effective action, and \(\vect{\chi} = (\chi_1, \chi_2, \chi_3, \ldots)\) is a tuple of internal variables (e.g., condensates).

To obtain the \textit{thermodynamic potential} \(\Omega(T,\mu)\), we must 
determine the values \(\bar{\boldsymbol{\chi}}(T,\mu)\) that minimize 
\(\Omega^\text{eff}\). These physical solutions, i.e.,\ the ones that 
actually describe the system’s realized state, satisfy the gap equations 
\begin{equation}\label{eq:condition}
\left[\frac{\partial \Omega^\text{eff}(T,\mu,\vect{\chi})}{\partial \chi_i}\right]_{\vect{\chi}=\bar{\vect{\chi}}} = 0 \quad \text{for all } i.
\end{equation}
The notation $[\cdots]_{\vect{\chi}=\bar{\vect{\chi}}}$ indicates a substitution of $\bar{\vect\chi}$ into $\vect\chi$. Substituting \(\bar{\vect{\chi}}(T,\mu)\) back into \(\Omega^\text{eff}\) yields
\begin{equation}\label{effactiodef}
\Omega(T,\mu) \equiv \Omega^\text{eff}(T,\mu,\bar{\vect{\chi}}(T,\mu)),
\end{equation}
which depends only on the external parameters \((T,\mu)\). The pressure is then given by \(P(T,\mu) = -\Omega(T,\mu)\).

Derivatives of \(\Omega(T,\mu)\) with respect to \(T\) and \(\mu\) are thus related to derivatives of \(\Omega^\text{eff}(T,\mu,\vect{\chi})\) evaluated at \(\bar{\vect{\chi}}(T,\mu)\). However, higher-order derivatives pose a challenge: they involve implicit dependencies on how derivatives of \(\bar{\vect{\chi}}(T,\mu)\) change with \(T\) and \(\mu\). Directly applying the chain rule leads to expressions with such dependencies. Derivatives of the internal parameters are not available symbolically.

To overcome this issue, we employ a \emph{Jacobian-based method}~\cite{Cox1965}. This approach provides a general framework for computing derivatives under constraints. If we consider a derivative of \(\Omega\) with respect to some function \(\mathcal{H}(T,\mu)\) at fixed \(\mathcal{G}(T,\mu)\), we can write
\begin{eqnarray}\label{eq;JacobianOmega}
\frac{\partial\Omega(T,\mu)}{\partial \mathcal{H}(T,\mu)}\bigg|_{\mathcal{G}(T,\mu)} 
&=& \frac{d(\Omega,\mathcal{G})/d(T,\mu)}{d(\mathcal{H},\mathcal{G})/d(T,\mu)} \nonumber \\
&=& \frac{
\left|\begin{array}{cc}
\partial_\mu \Omega & \partial_T \Omega \\
\partial_\mu \mathcal{G} & \partial_T \mathcal{G}
\end{array}\right|}{
\left|\begin{array}{cc}
\partial_\mu \mathcal{H} & \partial_T \mathcal{H} \\
\partial_\mu \mathcal{G} & \partial_T \mathcal{G}
\end{array}\right|}
,
\end{eqnarray}
where the vertical bars denote determinants and the subscripts indicate partial derivatives.
 The key advantage of this relation is that we can always express such derivatives as \(\mu\) and \(T\) derivatives of thermodynamic quantities such as $s$, $n$ and their derivatives. These quantities are the expansion coefficients in equation \eqref{eq:pressure_expansion_coefficients}. 
 
 One common application of this method is obtaining an expression for the speed of sound. The speed of sound squared is defined as 
\begin{equation}\label{eq:cstdeff}
c_s^2 = \left.\frac{\partial P}{\partial \epsilon}\right|_{s/n},
\end{equation}
where the derivative is taken at a fixed entropy-per-particle ratio \(s/n\). In a grand canonical ensemble, $P$, $\epsilon$, $s$ and $n$ are functions of $T$ and $\mu$. In most cases, one cannot express $P$, $\epsilon$, $s$ and $n$ as pure functions of each other without $T$ and $\mu$ dependence to be able to easily take the derivative in \eqref{eq:cstdeff}. By applying the Jacobian formalism, we convert this constrained derivative into a combination of partial derivatives of \(P\) and \(\epsilon\) with respect to \(T\) and \(\mu\) (see, for example, Ref.~\cite{shao2023speedsoundqcdmatter}). Using the structure of \eqref{eq;JacobianOmega}, Eq.~\eqref{eq:cstdeff} gives
\begin{eqnarray}\label{eq:cstexp}
    c_s^2 &=& \frac{1}{T s + \mu n} \;\; \frac{2 n s \frac{\partial s}{\partial \mu}|_T - s^2 \frac{\partial n}{\partial \mu}|_T - n^2 \frac{\partial s}{\partial T}|_\mu}{\left(\frac{\partial s}{\partial \mu}|_T\right)^2 - \frac{\partial s}{\partial T}|_\mu \frac{\partial n}{\partial \mu}|_T}.
\end{eqnarray}
In particular, the equation for \(c_s^2\) involves only partial derivatives of 
the thermodynamic potential with respect to \(\mu\) and \(T\), specifically 
\(\partial s/\partial \mu\), \(\partial n/\partial \mu\), and 
\(\partial s/\partial T\). These derivatives are precisely the higher-order 
pressure expansion coefficients. For instance,
\begin{align*}
c_{0,2} &= \frac{\partial^2 P}{\partial \mu^2}\bigg|_{T} =-\frac{\partial^2 \Omega}{\partial \mu^2}\bigg|_{T} = \frac{\partial n}{\partial \mu}\bigg|_{T}, \nonumber\\
c_{2,0} &= \frac{\partial^2 P}{\partial T^2}\bigg|_{\mu} =-\frac{\partial^2 \Omega}{\partial T^2} \bigg|_{\mu}=\frac{\partial s}{\partial T}\bigg|_{\mu}, \nonumber\\
c_{1,1} &= \frac{\partial^2 P}{\partial T \partial \mu} = -\frac{\partial^2 \Omega}{\partial T\partial \mu} =\frac{\partial s}{\partial \mu}\bigg|_{T} = \frac{\partial n}{\partial T}\bigg|_{\mu}.
\end{align*}
Since $c_s^2$ has a numeric value of order one, the subtraction operations in Eq.~\eqref{eq:cstexp} requires a high level of numerical precision to avoid significant digit loss. As a result, calculating these coefficients through numerical differentiation often introduces substantial errors, primarily due to the delicate cancellations in the numerator and denominator of Eq.~\eqref{eq:cstexp}.

In a finite-temperature QFT, the first-order coefficients $c_{1,0}$ (entropy) and $c_{0,1}$ (number density) are straightforward to obtain, as symbolic expressions for these quantities follow directly from the using a chain rule in substituting Eq.~\eqref{effactiodef} into Eq.~\eqref{eq:c01} (for more details, see Eq.~\eqref{eq:appfirstchain} in Appendix \ref{app:alternative})
\begin{align}\label{eq:sandn}
s(T,\mu) &= -\frac{\partial \Omega(T,\mu)}{\partial T} = - \left[\frac{\partial \Omega^\text{eff}(T,\mu,\vect{\chi})}{\partial T}\bigg|_{\mu,\vect\chi}\right]_{\vect{\chi}=\bar{\vect{\chi}}}, \\
n(T,\mu) &= -\frac{\partial \Omega(T,\mu)}{\partial \mu} = -\left[\frac{\partial \Omega^\text{eff}(T,\mu,\vect{\chi})}{\partial \mu}\bigg|_{T,\vect\chi}\right]_{\vect{\chi}=\bar{\vect{\chi}}}.
\end{align}
The notation $\big[\cdots\big|_{\vect\chi}\;\big]_{\vect{\chi}=\bar{\vect{\chi}}}$ indicates that, after performing the differentiation at fixed $\vect{\chi}$, we insert the physical solution $\bar{\vect{\chi}}(T,\mu)$.

Higher-order derivatives, however, are not as straightforward. For example, to compute $c_{2,0}=\partial^2 P/\partial T^2|_\mu$, one might attempt
\begin{align}\label{eq:higherchain}
\frac{\partial s(T,\mu)}{\partial T} &=- \frac{\partial}{\partial T}\left(\frac{\partial \Omega^\text{eff}(T,\mu,\bar{\vect{\chi}}(T,\mu))}{\partial T}\bigg|_{\mu}\right)\bigg|_{\mu}\nonumber\\[6pt]
&= -\left[\frac{\partial^2 \Omega^\text{eff}(T,\mu,\vect{\chi})}{\partial T^2}\bigg|_{\mu,\vect{\chi}}\right]_{\vect{\chi}=\bar{\vect{\chi}}}
\nonumber\\&- \left[\sum_i \frac{\partial \bar{\chi}_i}{\partial T}\bigg|_{\mu}\frac{\partial^2 \Omega^\text{eff}(T,\mu,\vect{\chi})}{\partial \chi_i \partial T}\bigg|_{\mu,\vect{\chi}\neq\chi_i}\right]_{\vect{\chi}=\bar{\vect{\chi}}}.
\end{align}
Here, the notation $\big|_{\vect{\chi}\neq \chi_i}$ means that when taking the partial derivative with respect to a specific condensate $\chi_i$, all other condensates in the set $\vect{\chi}$ are held fixed.
This expression introduces terms like $\partial \bar{\chi}_i/\partial T$, which are generally difficult to determine symbolically. Similar complications arise for $c_{1,1}$ and $c_{0,2}$.

Although alternative methods exist (e.g., as outlined in Appendix~A) to provide a symbolic expression for derivatives of $\bar{\vect\chi}$, they quickly become involved. Our goal is to present a more elegant and easily implementable solution. In the following section, we introduce a Jacobian-based method as in Eq.~\eqref{eq;JacobianOmega}, enabling the computation of arbitrary higher-order coefficients without explicitly solving for the internal parameter dependencies. This approach provides a unified symbolic framework that avoids the shortcomings of direct numerical differentiation.

\section{Calculation of Higher Order Expansion Coefficients}\label{sec:3}

In this section, we employ the Jacobian method to derive symbolic expressions\footnote{By “symbolic'', we mean that the final formula is presented in an expression involving only derivatives of the effective potential, which are accessible symbolically. The only numerical step required is the insertion of the solutions to the gap equations into this final symbolic expression.
} for higher-order thermodynamic derivatives. This approach allows us to bypass the need to explicitly determine how derivatives of internal parameters depend on external variables, a necessity that complicates direct chain rule applications (see for example Eq.~\eqref{eq:higherchain}). We first confirm that, for the first derivatives, our symbolic method reproduces the known expressions. We then extend it to higher-order derivatives, where its advantages become particularly evident.

To illustrate the procedure, consider a system with a single internal variable \(\chi\). Starting with the first $T$-derivative, we have for \( c_{1,0} \)
\begin{align}\label{eq:c10temp}
c_{1,0} &= -\frac{\partial}{\partial T}\Omega(T,\mu)\bigg|_\mu \nonumber \\
&= -\frac{\partial}{\partial T}\bigg(\bigg[\Omega^\text{eff}(T,\mu,\chi)\bigg]_{\chi=\bar\chi(T,\mu)}\bigg)\bigg|_\mu.
\end{align}
Applying the chain rule at this point would reproduce Eq.~\eqref{eq:sandn} (details in Appendix \ref{app:alternative}). However, to proceed purely in a symbolic manner, we perform a Legendre transform of the effective potential from the variables $(T,\mu,\chi)$ to $(T,\mu,q\equiv\partial_\chi\Omega^\text{eff})$. Defining
\begin{align}
\Omega^\text{eff}_{\text{Transformed}}(T,\mu,q) &= -\Omega^\text{eff}(T,\mu,\tilde\chi(T,\mu,q)) \nonumber\\
&\quad+ q\,\partial_\chi\Omega^\text{eff}(T,\mu,\tilde\chi(T,\mu,q)),
\end{align}
where $\tilde\chi$ is determined by $q=\partial_\chi\Omega^\text{eff}|_{\chi=\tilde\chi}$. The relation between $\bar\chi$ and $\tilde\chi$ can be realized as $\bar\chi(T,\mu)=\tilde\chi(T,\mu,0)$. We find
\begin{eqnarray}
\Omega(T,\mu) &=& -\Omega^\text{eff}_{\text{Transformed}}(T,\mu,q=0) \nonumber\\
&=& \Omega^\text{eff}(T,\mu,\tilde\chi(T,\mu,0)).
\end{eqnarray}
Using this relation, we rewrite Eq.~\eqref{eq:c10temp} as
\begin{eqnarray}
c_{1,0} = -\frac{\partial}{\partial T}\bigg[\Omega^\text{eff}(T,\mu,\tilde\chi(T,\mu,q))\bigg]_{q=0}\bigg|_{\mu}.
\end{eqnarray}
Since $T,\mu,$ and $q$ are independent variables, we can first take the $T$-derivative at fixed $\mu$ and $q$, and only then impose $q=0$
\begin{eqnarray}
c_{1,0} = -\bigg[\frac{\partial}{\partial T}\Omega^\text{eff}(T,\mu,\tilde\chi(T,\mu,q))\bigg|_{\mu,q}\bigg]_{q=0}.
\end{eqnarray}
Setting $q=0$ corresponds to the physical condition $\partial_\chi\Omega^\text{eff}=0$, which defines $\bar{\chi}(T,\mu)$ in Eq.~\eqref{eq:condition}. Thus,
\begin{eqnarray}\label{c10full}
c_{1,0} = \left[-\frac{\partial}{\partial T}\Omega^\text{eff}(T,\mu,\chi)\bigg|_{\mu,\partial_{\chi}\Omega^\text{eff}}\right]_{\partial_{\chi}\Omega^\text{eff}=0,\chi=\bar\chi}.
\end{eqnarray}
Within the brackets, we have a derivative taken under the constraint that value of $\partial_{\chi}\Omega^\text{eff}$ is fixed. By setting $\partial_{\chi}\Omega^\text{eff}$ to zero after differentiation, we recover the physical solution.--- This corresponds to imposing the same condition as in Eq.~\eqref{eq:condition}, where the value of \( q \) is zero. This defines \(\bar{\chi}(T,\mu)\) as the physical solution.
Once we know $\bar{\chi}(T,\mu)$ (usually obtained numerically by an optimization procedure in Eq.~\eqref{eq:condition}), we insert it into the final symbolic expressions.

To generalize this procedure, we introduce an intermediate quantity,
\begin{equation}
\tilde{c}_{m,n}(T,\mu,\chi) \equiv \frac{\partial^{m+n}\Omega^\text{eff}(T,\mu,\chi)}{\partial T^m \partial \mu^n}\bigg|_{\partial_\chi\Omega^\text{eff}=0}.
\end{equation}
Here, the notation indicates that the derivative is taken at fixed $\partial_\chi\Omega^\text{eff}$, and only afterward do we set $\partial_\chi\Omega^\text{eff}=0$. Our assumption is that, given the symbolic expression for $\Omega^\text{eff}$, the quantities $\tilde{c}_{m,n}$ can likewise be derived in symbolic form. Specifically, as we will see in the following, from the Jacobian expansion of the above equation, all derivatives of the effective potential of the form $\frac{\partial^{m+n+l}\Omega^\text{eff}}{\partial T^m \partial \mu^n \partial \chi^l}$ are available symbolically.

The relationship between the effective potential $\Omega^\text{eff}$ and the thermodynamic potential $\Omega$ is just like the relationship between the symbolic coefficients $\tilde{c}_{m,n}$ and the pressure expansion coefficients $c_{m,n}$. Just as substituting the solutions of the gap equations into $\Omega^\text{eff}$ produces the thermodynamic potential $\Omega$, inserting $\chi=\bar{\chi}(T,\mu)$ into the symbolic expressions $\tilde{c}_{m,n}$ yields the pressure expansion coefficients $c_{m,n}$. Once we have the symbolic expressions for $\tilde{c}_{m,n}$, we obtain $c_{m,n}$ by inserting the physical solution $\chi=\bar{\chi}(T,\mu)$, just as we obtain $\Omega$ from $\Omega^\text{eff}$ in Eq.~\eqref{effactiodef}

\begin{equation}\label{eq:tildec}
c_{m,n}(T,\mu) = \big[\tilde{c}_{m,n}(T,\mu,\chi)\big]_{\chi=\bar{\chi}(T,\mu)}.
\end{equation}
By following this logic for the first derivatives, we reproduce Eq.~\eqref{eq:sandn}. Using the structure in Eq.~\eqref{eq;JacobianOmega}, the term inside the brackets in Eq.~\eqref{c10full} gives
\begin{align}\label{eq:firstTder}
\tilde c_{1,0}&=-\frac{\partial}{\partial T}\left(\Omega^\text{eff}\left(T,\mu,\chi\right)\right)\bigg|_{\mu,\partial_\chi\Omega^\text{eff}(T,\mu,\chi)=0}\nonumber\\
&=-\bigg[ \frac{d(\Omega^\text{eff},\mu,\partial_\chi\Omega^\text{eff})}{d(T,\mu,\partial_\chi\Omega^\text{eff})}\bigg]_{\partial_\chi\Omega^\text{eff}(T,\mu,\chi)=0}.
\end{align}
Inside the last brackets, we transform the numerator and denominator by dividing \( d(T,\mu,\chi) \) to obtain
\begin{align}{\label{eq:jacotder}}
\frac{\dfrac{d(\Omega^\text{eff},\mu,\partial_\chi\Omega^\text{eff})}{d(T,\mu,\chi)}}{\dfrac{d(T,\mu,\partial_\chi\Omega^\text{eff})}{d(T,\mu,\chi)}} = \frac{
\left|\begin{array}{cc}
\partial_T \Omega^\text{eff} & \partial_T (\partial_\chi \Omega^\text{eff}) \\[5pt]
\partial_\chi \Omega^\text{eff} & \partial_\chi (\partial_\chi \Omega^\text{eff})
\end{array}\right|
}{
\partial_\chi (\partial_\chi \Omega^\text{eff})
}.
\end{align}
In the above expressions, the chemical potential \(\mu\) appears both in the numerator and the denominator of the Jacobians. Due to the properties of Jacobians, we can drop the \(\mu\) dependence as we did in the third line. For further details on this simplification and the underlying notation, refer to Ref.~\cite{Cox1965}. Eq.~\eqref{eq:jacotder} represents the derivative under the constraint that $\partial_\chi\Omega^\text{eff}$ remains any fixed constant. To obtain $\tilde{c}_{1,0}$ from Eq.~\eqref{eq:firstTder}, we must impose the substitution $\partial_\chi\Omega^\text{eff}(T,\mu,\chi)=0$, that is, we set the lower-left entry of the determinant to zero. To get the expansion coefficient $c_{1,0}$, what remains is the insertion in Eq.~\eqref{eq:tildec}, which imposes setting $\vect\chi$ to $\bar{\vect\chi}$ and we get
\begin{align}\label{eq:structure}
c_{1,0}=-\frac{\partial}{\partial T}\Omega\left(T,\mu\right) &= -\bigg[\left(\frac{
\left|\begin{array}{cc}
\partial_T \Omega^\text{eff} & \partial_T (\partial_\chi \Omega^\text{eff}) \\[5pt]
0 & \partial_\chi (\partial_\chi \Omega^\text{eff})
\end{array}\right|
}{
\partial_\chi (\partial_\chi \Omega^\text{eff})
}\right)\bigg]_{\chi = \bar{\chi}} \nonumber \\
&=- \bigg[\left(\frac{\partial}{\partial T}\Omega^\text{eff}(T,\mu,\chi)\bigg|_{\mu,\chi}\right)\bigg]_{\chi = \bar{\chi}},
\end{align}
which is the same expression in Eq.~\eqref{eq:sandn}. We obtain the same expression for number density \( n \) by using this method for the first chemical potential derivative of the pressure.
By applying the same reasoning to higher-order derivatives, as we will demonstrate shortly, we circumvent the complexity of explicitly computing internal parameter derivatives such as $\partial \bar{\chi}_i/\partial T$ or $\partial \bar{\chi}_i/\partial \mu$, which appear in Eq.~\eqref{eq:higherchain}. As an example, we will calculate the second $T$-derivative $c_{2,0}$ by applying the same procedure used for the first $T$-derivative to the expression in Eq.~\eqref{eq:jacotder}. We get

\begin{align}\label{eq:stjac}
c_{2,0} &=- \frac{\partial}{\partial T}\left(\frac{\partial}{\partial T}\Omega(T,\mu)\right)\nonumber \\
&=-\frac{d(\partial_T\Omega^\text{eff},\mu,\partial\chi\Omega^\text{eff})/d(T,\mu,\chi)}{d(T,\mu,\partial_\chi\Omega^\text{eff})/d(T,\mu,\chi)}
\nonumber\\
&=-\bigg[ \left(\frac{
\left|\begin{array}{cc}
\partial_T (\partial_T\Omega^\text{eff}) & \partial_T (\partial_\chi \Omega^\text{eff}) \\[5pt]
\partial_\chi (\partial_T \Omega^\text{eff}) & \partial_\chi (\partial_\chi \Omega^\text{eff})
\end{array}\right|
}{
\partial_\chi (\partial_\chi \Omega^\text{eff})
}\right)\bigg]_{\chi = \bar{\chi}} \nonumber\\
&=- \bigg[\left(
\frac{\partial^2 \Omega^\text{eff}(T,\mu, \chi)}{\partial T^2} 
- \frac{\left(\dfrac{\partial^2 \Omega^\text{eff}(T,\mu, \chi)}{\partial T \partial \chi}\right)^2}{\dfrac{\partial^2 \Omega^\text{eff}(T,\mu, \chi)}{\partial \chi^2}}
\right) \bigg]_{\chi = \bar{\chi}}.
\end{align}
For $c_{1,1}=\partial s/\partial \mu\big|_T$ and $c_{0,2}=\partial n/\partial \mu\big|_T$, following the same steps gives
\begin{align}
c_{1,1}=&\frac{\partial s}{\partial \mu}\bigg|_T = -\bigg[\!
\frac{\partial^2 \Omega^\text{eff}}{\partial T \partial \mu} 
- \frac{(\partial_\mu \partial_\chi \Omega^\text{eff})(\partial_T \partial_\chi \Omega^\text{eff})}{\partial_\chi^2 \Omega^\text{eff}}
\bigg]_{\chi = \bar{\chi}},\label{eq:c11f}\\[6pt]
c_{0,2}=&\frac{\partial n}{\partial \mu}\bigg|_T = -\bigg[\!
\frac{\partial^2 \Omega^\text{eff}}{\partial \mu^2}
- \frac{(\partial_\mu \partial_\chi \Omega^\text{eff})^2}{\partial_\chi^2 \Omega^\text{eff}}
\bigg]_{\chi = \bar{\chi}}.\label{eq:c02f}
\end{align}
In all these expressions, the term inside the brackets can be expressed entirely in symbolic form. Unlike Eq.~\eqref{eq:higherchain}, this approach avoids any direct computation of internal parameter derivatives. The key insight is that, by following this procedure, all higher-order derivatives can be expressed solely in terms of derivatives of $\Omega^\text{eff}$, which provides a symbolic expression as intended.

In Eqs.~\eqref{eq:stjac},\eqref{eq:c11f} and \eqref{eq:c02f}, the expression for the second derivative consists of two terms. The first term matches the first term in the chain rule result from Eq.~\eqref{eq:higherchain}, and can be interpreted as what we would obtain if we naively ignored the implicit dependence of the internal parameters on $T$ and $\mu$. More generally, for any higher-order derivative, we define the \emph{naive derivative} as the portion of the pressure expansion coefficient that remains when the implicit dependence of $\chi$ on the external variables is neglected

\begin{align}\label{eq:naivederivative}
    \frac{\partial^{m+n}}{\partial T^m\partial\mu^n}\Omega
    = \underbrace{\bigg[
    \frac{\partial^{m+n}}{\partial T^m\partial\mu^n}\Omega^\text{eff}
    \bigg]_{\chi=\bar{\chi}}}_{\text{naive derivative}}
    + \text{correction terms}.
\end{align}
The full expression includes correction terms beyond the naive derivative to properly account for the constraints involved in taking the derivative. These corrections become crucial in regions where $\bar{\chi}(T,\mu)$ varies rapidly, such as near phase transitions, ensuring that we capture the correct physical behavior.

To illustrate this further, consider a system exhibiting a first-order phase transition at some critical chemical potential $\mu_c$. If the order parameter is given by an internal parameter $\chi$, it will jump discontinuously from $\chi_a$ to $\chi_b$ at $\mu=\mu_c$. In this scenario, even the naive derivative captures the discontinuity associated with the first-order phase transition, since inserting the appropriate $\chi$ values (either $\chi_a$ or $\chi_b$) into the expression reflects the sudden change. However, in the case of a second-order or crossover phase transition—where the order parameter changes continuously—the naive derivative alone will vary smoothly and fail to reflect the underlying critical behavior. By including the correction terms, our symbolic method can accurately reproduce the characteristic signatures of these continuous transitions. This is clearly demonstrated in the examples presented in the  Sec.~\ref{sec:examp}.

Using the second-order pressure coefficients from Eqs.~\eqref{eq:stjac}, \eqref{eq:c11f}, and \eqref{eq:c02f}, we can insert them into Eq.~\eqref{eq:cstexp} to derive a fully symbolic expression for the speed of sound squared, $c_s^2(T,\mu)$. These results also enable us to compute other thermodynamic properties that depend on second-order derivatives of the thermodynamic potential, such as the heat capacity $c_v$ and the particle number susceptibility $\chi_p$. Furthermore, by systematically extending this procedure, one can obtain higher-order pressure coefficients of arbitrary order, providing a consistent and transparent symbolic framework for all required thermodynamic derivatives. For example, third-order coefficients can be used to calculate shock adiabats~\cite{PhysRevD.40.2903,Rischke_1996} or to investigate scenarios similar to those discussed in Ref.~\cite{Mroczek:2024sfp}.

\subsection*{General Case of Multiple Internal Variables}

Having established the procedure for a single internal parameter, we now generalize it to the case where multiple internal parameters $\vect{\chi}=(\chi_i)$ are present. The logic follows the same principles discussed previously, but the algebra becomes more involved as the dimensionality of the internal parameter space increases.

We begin by revisiting the calculation of $c_{1,0}$, now in the presence of multiple $\chi_i$. Starting from the structure analogous to Eq.~\eqref{eq:structure}, we have
\begin{eqnarray}\label{structure3}
c_{1,0}&=&-\bigg[\frac{\partial}{\partial T}\Big(\Omega^\text{eff}(T,\mu,\vect{\chi})\big|_{\{\partial_{\chi_i}\Omega^\text{eff}\}}\Big)\bigg]_{\vect{\chi}=\bar{\vect{\chi}}}\\
&=&-\bigg[\frac{d(\Omega^\text{eff},\mu,\partial_{\chi_1}\Omega^\text{eff},\partial_{\chi_2}\Omega^\text{eff},\partial_{\chi_3}\Omega^\text{eff},\ldots)}{d(T,\mu,\partial_{\chi_1}\Omega^\text{eff},\partial_{\chi_2}\Omega^\text{eff},\partial_{\chi_3}\Omega^\text{eff},\ldots)}\bigg]_{\vect{\chi}=\bar{\vect{\chi}}}\nonumber\\
&=&-\bigg[\frac{\frac{d(\Omega^\text{eff},\partial_{\chi_1}\Omega^\text{eff},\partial_{\chi_2}\Omega^\text{eff},\partial_{\chi_3}\Omega^\text{eff},\ldots)}{d(T,\chi_1,\chi_2,\chi_3,\ldots)}}{\frac{d(T,\partial_{\chi_1}\Omega^\text{eff},\partial_{\chi_2}\Omega^\text{eff},\partial_{\chi_3}\Omega^\text{eff},\ldots)}{d(T,\chi_1,\chi_2,\chi_3,\ldots)}}\bigg]_{\vect{\chi}=\bar{\vect{\chi}}}\,.\nonumber
\end{eqnarray}
Here in the second line $\mu$ is dropped as before in Eq.~\eqref{eq:jacotder}. In the last line, we have transformed the Jacobian by dividing both the numerator and denominator by $d(T,\chi_1,\chi_2,\chi_3,\ldots)$. To simplify notation, we introduce
\begin{align}
J\!\left(\frac{\mathcal{F},\partial_{\vect{\chi}}\Omega^\text{eff}}{y,\vect{\chi}}\right) &\equiv \frac{d(\mathcal{F},\partial_{\chi_1}\Omega^\text{eff},\partial_{\chi_2}\Omega^\text{eff},\partial_{\chi_3}\Omega^\text{eff},\ldots)}{d(y,\chi_1,\chi_2,\chi_3,\ldots)} \nonumber\\\nonumber\\
&\mkern-90mu=
\begin{vmatrix}
\partial_y\mathcal{F} & \partial_y \partial_{\chi_1}\Omega^\text{eff} & \partial_y \partial_{\chi_2}\Omega^\text{eff} & \partial_y \partial_{\chi_3}\Omega^\text{eff} & \cdots \\[6pt]
\partial_{\chi_1}\mathcal{F} & \partial_{\chi_1}^2\Omega^\text{eff} & \partial_{\chi_1}\partial_{\chi_2}\Omega^\text{eff} & \partial_{\chi_1}\partial_{\chi_3}\Omega^\text{eff} & \cdots \\[6pt]
\partial_{\chi_2}\mathcal{F} & \partial_{\chi_2}\partial_{\chi_1}\Omega^\text{eff} & \partial_{\chi_2}^2\Omega^\text{eff} & \partial_{\chi_2}\partial_{\chi_3}\Omega^\text{eff} & \cdots \\[6pt]
\partial_{\chi_3}\mathcal{F} & \partial_{\chi_3}\partial_{\chi_1}\Omega^\text{eff} & \partial_{\chi_3}\partial_{\chi_2}\Omega^\text{eff} & \partial_{\chi_3}^2\Omega^\text{eff} & \cdots \\[6pt]
\vdots & \vdots & \vdots & \vdots & \ddots
\end{vmatrix},\label{eq:numstruct} \\[6pt]
J\!\left(\frac{\partial_{\vect{\chi}}\Omega^\text{eff}}{\vect{\chi}}\right) &\equiv \frac{d(\partial_{\chi_1}\Omega^\text{eff},\partial_{\chi_2}\Omega^\text{eff},\partial_{\chi_3}\Omega^\text{eff},\ldots)}{d(\chi_1,\chi_2,\chi_3,\ldots)} \nonumber\\\nonumber\\
&\mkern-60mu=
\begin{vmatrix}
\partial_{\chi_1}^2\Omega^\text{eff} & \partial_{\chi_1}\partial_{\chi_2}\Omega^\text{eff} & \partial_{\chi_1}\partial_{\chi_3}\Omega^\text{eff} & \cdots \\[6pt]
\partial_{\chi_2}\partial_{\chi_1}\Omega^\text{eff} & \partial_{\chi_2}^2\Omega^\text{eff} & \partial_{\chi_2}\partial_{\chi_3}\Omega^\text{eff} & \cdots \\[6pt]
\partial_{\chi_3}\partial_{\chi_1}\Omega^\text{eff} & \partial_{\chi_3}\partial_{\chi_2}\Omega^\text{eff} & \partial_{\chi_3}^2\Omega^\text{eff} & \cdots \\[6pt]
\vdots & \vdots & \vdots & \ddots
\end{vmatrix},
\label{eq:denstruct}
\end{align}
where $y$ is either $T$ or $\mu$.
Using these definitions and applying the condition $\partial_{\chi_i}\Omega^\text{eff}=0$ for all $i$, we obtain
\begin{subequations}\label{eq:firstst}
\begin{eqnarray}
c_{1,0}&=&-\partial_T \Omega(T,\mu) = -\bigg[\frac{J\!\left(\frac{\Omega^\text{eff}, \partial_{\vect{\chi}} \Omega^\text{eff}}{T, \vect{\chi}}\right)}{J\!\left(\frac{\partial_{\vect{\chi}} \Omega^\text{eff}}{\vect{\chi}}\right)}\bigg]_{\vect{\chi} = \bar{\vect{\chi}}}\nonumber\\
&=&\bigg[\partial_T \Omega^\text{eff}(T,\mu,\vect{\chi})\bigg]_{\vect{\chi} = \bar{\vect{\chi}}}, \\[6pt]
c_{0,1}&=&-\partial_\mu \Omega(T,\mu) 
= -\bigg[\frac{J\!\left(\frac{\Omega^\text{eff}, \partial_{\vect{\chi}} \Omega^\text{eff}}{\mu, \vect{\chi}}\right)}{J\!\left(\frac{\partial_{\vect{\chi}} \Omega^\text{eff}}{\vect{\chi}}\right)}\bigg]_{\vect{\chi} = \bar{\vect{\chi}}}\nonumber\\
&=&-\bigg[\partial_\mu \Omega^\text{eff}(T,\mu,\vect{\chi})\bigg]_{\vect{\chi} = \bar{\vect{\chi}}}.
\end{eqnarray}
\end{subequations}
With this structure in place, one can systematically compute higher-order pressure expansion coefficients. By repeatedly inserting known lower-order coefficients into the determinant expressions, we build up to arbitrary $(m,n)$ derivatives. Defining $\tilde{c}_{m,n}$ as in Eq.~\eqref{eq:tildec},
\begin{eqnarray}\label{eq:substi}
c_{m,n}(T,\mu) \equiv \bigg[\tilde{c}_{m,n}(T,\mu,\vect{\chi})\bigg]_{\vect{\chi} = \bar{\vect{\chi}}},
\end{eqnarray}
we arrive at recursive relations for $(m+n)$-th order derivatives
\begin{subequations}\label{eq:genstruct}
\begin{eqnarray}
c_{m,n}(T,\mu)&=&-\frac{\partial^{m+n}\Omega(T,\mu)}{\partial T^m\partial \mu^n}\nonumber\\[6pt]
&=&\bigg[\frac{J\!\left(\frac{\tilde{c}_{m-1,n},\partial_{\vect{\chi}}\Omega^\text{eff}}{T,\vect{\chi}}\right)}{J\!\left(\frac{\partial_{\vect{\chi}}\Omega^\text{eff}}{\vect{\chi}}\right)}\bigg]_{\chi=\bar{\chi}},
\end{eqnarray}
or equivalently
\begin{eqnarray}
c_{m,n}(T,\mu)&=&\bigg[\frac{J\!\left(\frac{\tilde{c}_{m,n-1},\partial_{\vect{\chi}}\Omega^\text{eff}}{\mu,\vect{\chi}}\right)}{J\!\left(\frac{\partial_{\vect{\chi}}\Omega^\text{eff}}{\vect{\chi}}\right)}\bigg]_{\chi=\bar{\chi}}.
\end{eqnarray}
\end{subequations}
From these formulas, it is evident that each $c_{m,n}$ can be expressed entirely in terms of partial derivatives of $\Omega^\text{eff}$ and the Jacobian structures encoding the constraints. As noted earlier, a naive derivative contribution always appears, corresponding to the direct $(m+n)$-th derivative of $\Omega^\text{eff}$ without any constraint corrections (see Eq.~\eqref{eq:naivederivative}). This becomes clear when we expand the numerator determinant by its minors: the first term in this expansion is precisely the naive derivative multiplied by the determinant that appears in the numerator. Hence, the naive derivative is always present as a baseline component of the full expression.

In practice, as the number of internal parameters increases, the complexity of these Jacobian expressions grows significantly, making the process algebraically demanding. Following Eq.~\eqref{eq:denstruct}, for \( N \) internal variables, the denominator involves a fixed number of independent terms: \( N(N+1)/2 \). These terms need to be defined only once, as they remain the same and repeat at each step in Eq.~\eqref{eq:genstruct}. For the numerator, however, additional unique expressions must be handled at each step, accounting for all combinations of derivatives with respect to \( T, \mu \), and \( \chi \) up to the relevant order (compare Eq.~\eqref{eq:numstruct} to Eq.~\eqref{eq:denstruct}).

Specifically, for the first derivatives of pressure (\(c_{1,0}\) and \(c_{0,1}\)), one must handle an additional \(N+2\) unique expressions. These correspond to the derivatives of the effective potential with respect to \(T\), \(\mu\), and the \(N\) internal variables, which are determined from the gap equations. For the second derivatives of pressure (\(c_{2,0}\), \(c_{1,1}\), and \(c_{0,2}\)), an additional \(\frac{(N+2)(N+3)}{2}\) expressions are required. These include all second-order combinations: \(\partial_T^2 \Omega^\text{eff}\), \(\partial_T \partial_\mu \Omega^\text{eff}\), \(\partial^2_\mu \Omega^\text{eff}\), \(\partial_T \partial_{\chi_i} \Omega^\text{eff}\), \(\partial_\mu \partial_{\chi_i} \Omega^\text{eff}\), and \(\partial_{\chi_i} \partial_{\chi_j} \Omega^\text{eff}\). For the third derivatives of pressure, an additional \(\frac{(N+2)(N+3)(N+4)}{6}\) expressions must be defined, which account for all third-order derivative combinations of the effective potential. For \(N = 1\), \(N = 2\), and \(N = 3\), the additional unique expressions required are \(3\), \(4\), and \(5\) for the first derivatives; \(6\), \(10\), and \(15\) for the second derivatives; and \(10\), \(20\), and \(35\) for the third derivatives, respectively. This is showcased in Appendix~\ref{app:expansion}, where explicit calculations for $N=1$ and  $N=2$ are detailed. In general, the total number of unique expressions required up to an arbitrary order \( L \) is given by summing the contributions for all derivatives from \( k=1 \) to \( L \): \(  \sum_{k=1}^L \binom{N+2+k-1}{k}=\binom{N + L + 2}{L} - 1\sim \frac{(N+L+2)^L}{L!} \). This clearly shows the growth in complexity as \( N \) and \( L \) increase.

To overcome these challenges, a practical and efficient solution is to employ Automatic Differentiation (AD) for evaluating the derivatives that appear in the Jacobians of Eq.~\eqref{eq:genstruct}. AD is a computational technique that systematically and efficiently computes derivatives of functions by breaking down complex expressions into a sequence of elementary operations and applying the chain rule automatically. Unlike symbolic differentiation, which manipulates expressions analytically, or numerical differentiation using finite differences, AD provides exact derivative values up to machine precision without the cost of analytic manipulation. This makes AD particularly well-suited for handling the intricate derivative structures that arise in multi-parameter models.  Notably, the utility of AD has been demonstrated in new computational frameworks such as the \texttt{DiFfRG} framework (Ref.~\cite{sattler2024diffrgdiscretisationframeworkfunctional}), which integrates AD for solving FRG flow equations, and in the work of Ref.~\cite{ihssen2023quantitativeprecisionqcdlarge}, where AD is combined with advanced discretization techniques to enhance precision in resolving QCD phase diagrams.

However, it is important to note that for our method of calculations in this paper, applying AD directly at the top level to $\Omega^\text{eff}$ \underline{\textbf{does not}} automatically incorporate the constraints on internal parameters or their implicit dependence on $T$ and $\mu$ (see Ref.~\cite{schaefer2009qcdthermodynamicseffectivemodels} and Ref.~\cite{Wagner_2010} for a method on handling implicit dependencies in AD frameworks for calculating Taylor expansion coefficients of thermodynamic potential). Instead, AD can be utilized to compute the expressions obtained after applying our proposed method. By doing so, we ensure a fully systematic and automatable procedure for calculating higher-order pressure coefficients in more complex mean-field models, while significantly reducing the algebraic complexity of these computations.

\begin{figure*}[t]
    \begin{minipage}[t]{0.45\textwidth}
        \centering
        \includegraphics[height=0.23\textheight]{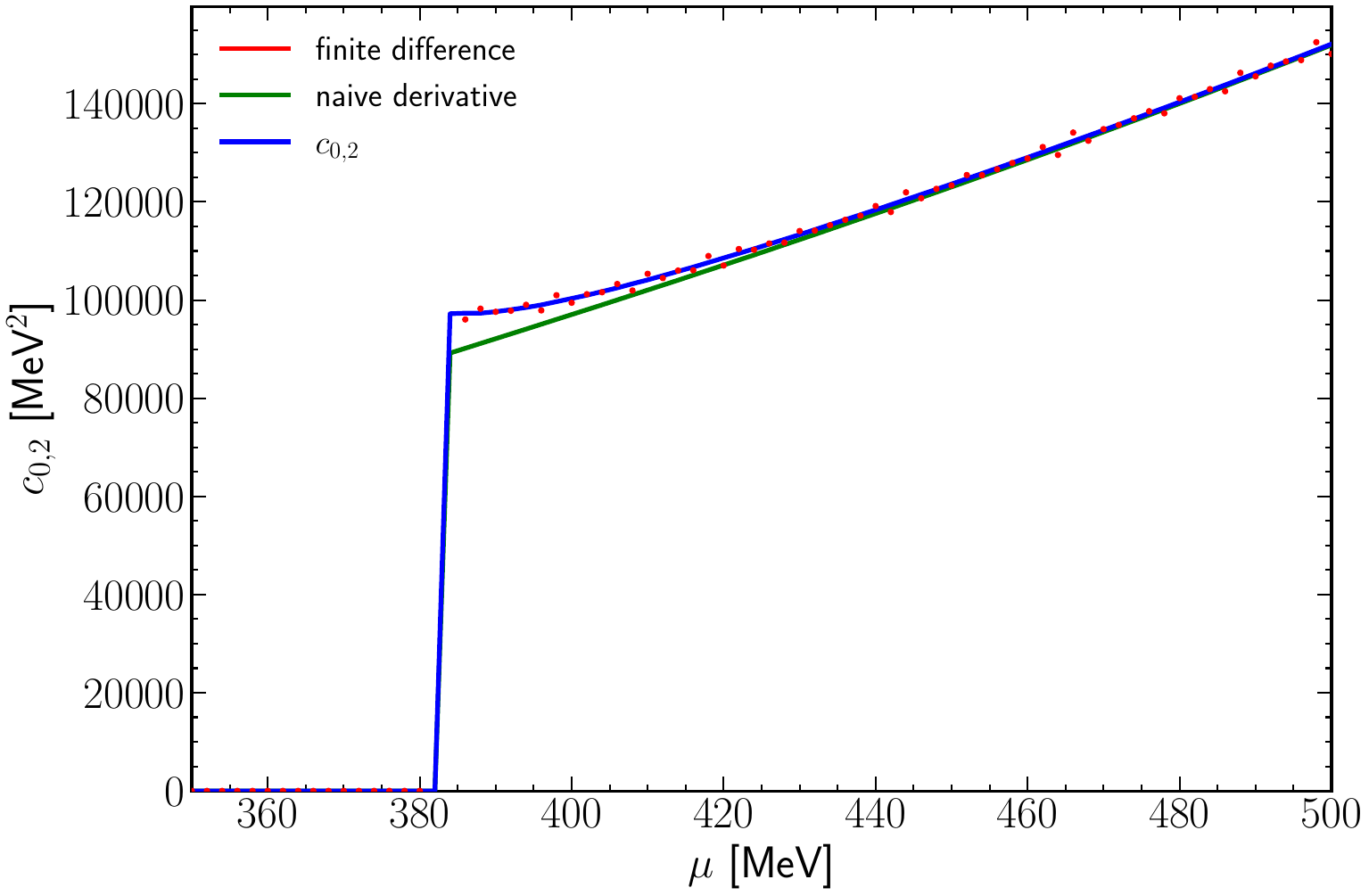}
    \end{minipage}
    \hfill
    \begin{minipage}[t]{0.45\textwidth}
        \centering
        \includegraphics[height=0.23\textheight]{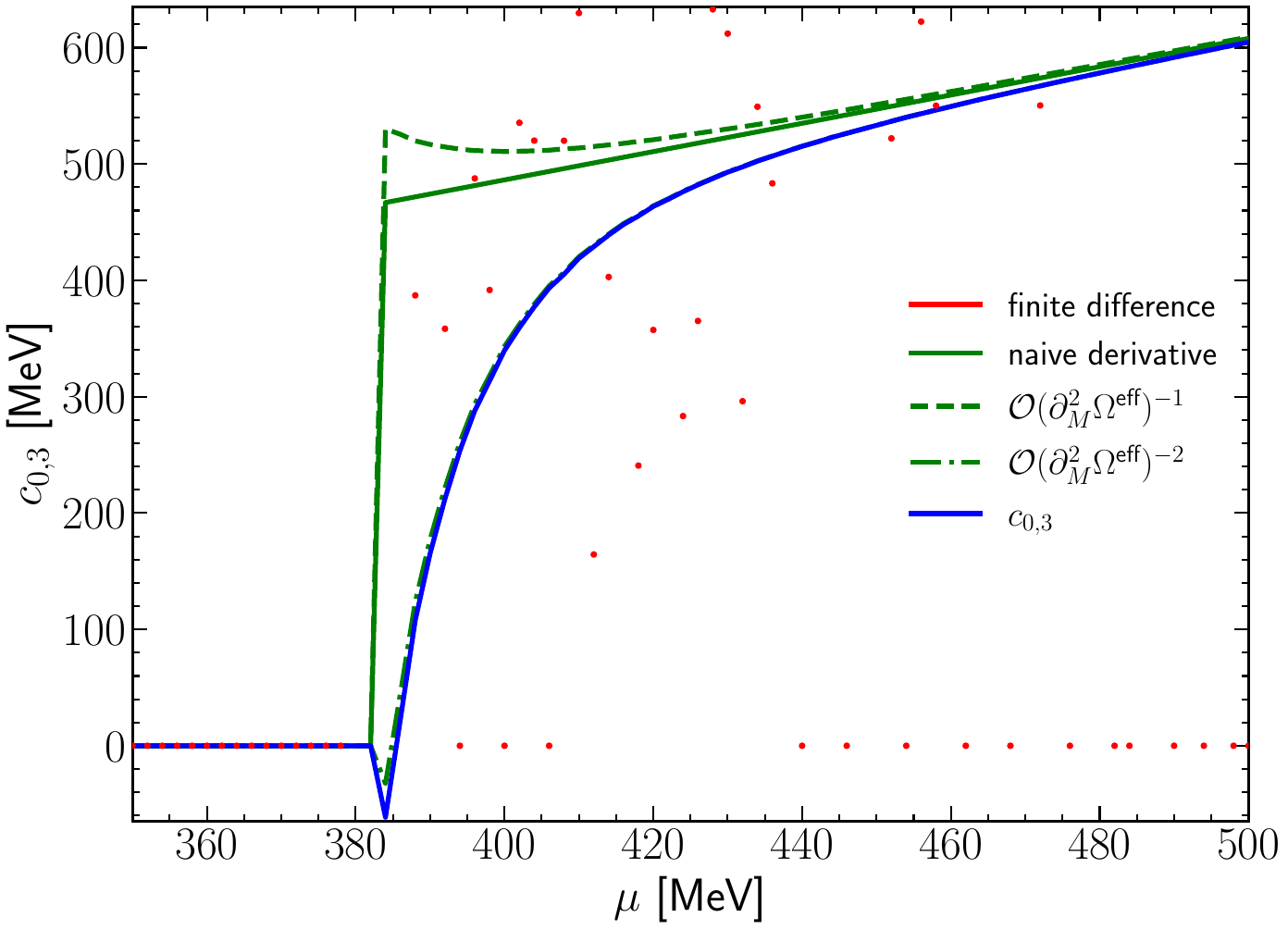}
    \end{minipage}
    \caption{Second- and third-order $\mu$-derivatives of the pressure at $T = 0.1 \,\text{MeV}$:
    the chemical potential expansion coefficients $c_{0,2} = \partial^2 P/\partial \mu^2$ (left) and $c_{0,3} = \partial^3 P/\partial \mu^3$ (right) as functions of $\mu$. The red dots denote numerical finite difference results, which show large fluctuations for $c_{0,2}$ and thus yield unreliable data. In contrast, the symbolic derivatives (blue lines) provide stable and accurate results for both $c_{0,2}$ and $c_{0,3}$. The green curves illustrate approximations to the symbolic result by truncating the expansion in powers of $(\partial_M^2\Omega^\text{eff})^{-1}$, as indicated in Eq.~\eqref{eq:pressure_derivatives_tilde}.}
    \label{fig:tempcoeff}
\end{figure*}

\begin{figure*}[t]
    \begin{minipage}[t]{0.45\textwidth}
        \centering
        \includegraphics[height=0.23\textheight]{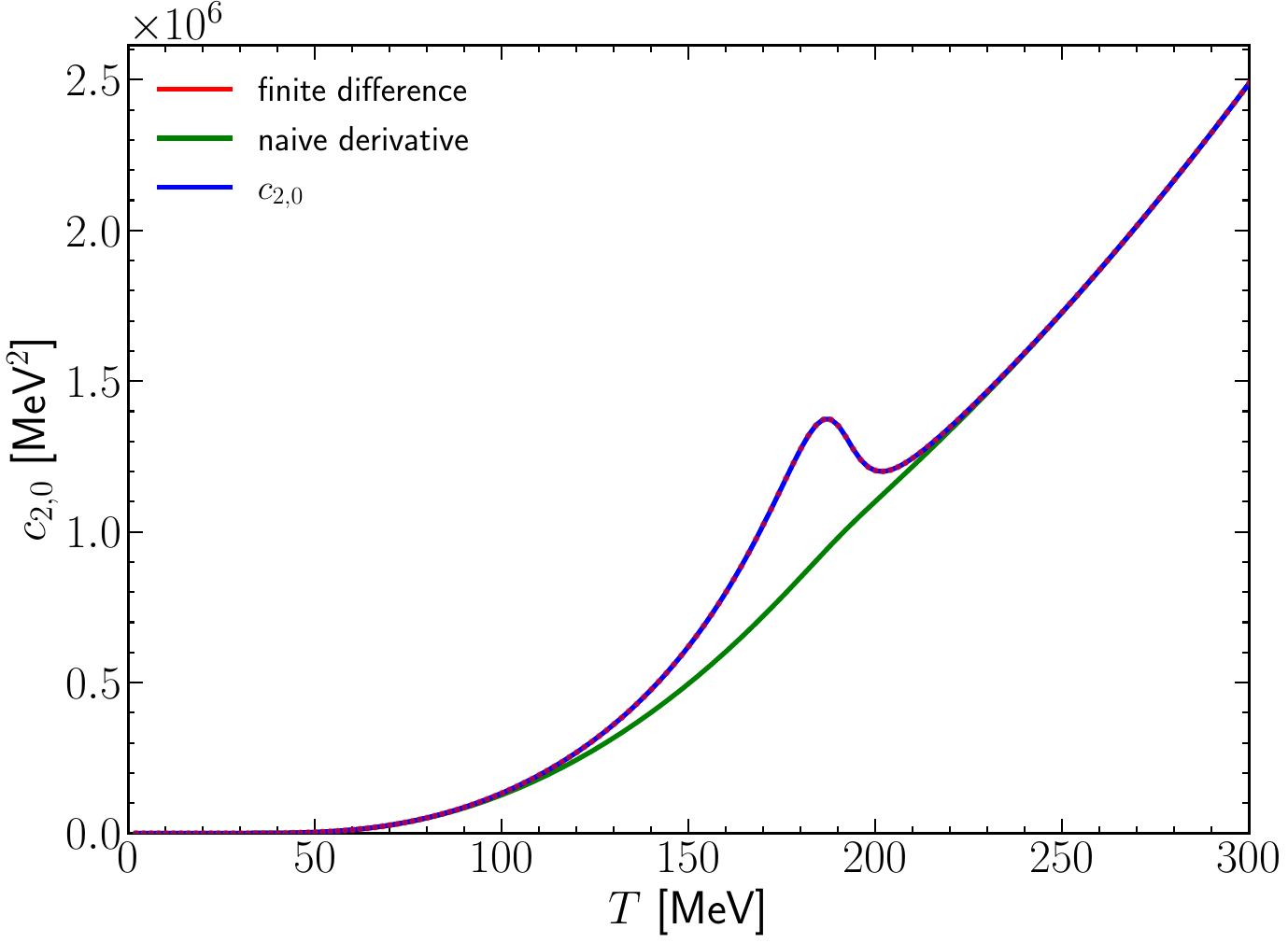}
    \end{minipage}
    \hfill
    \begin{minipage}[t]{0.45\textwidth}
        \centering
        \includegraphics[height=0.23\textheight]{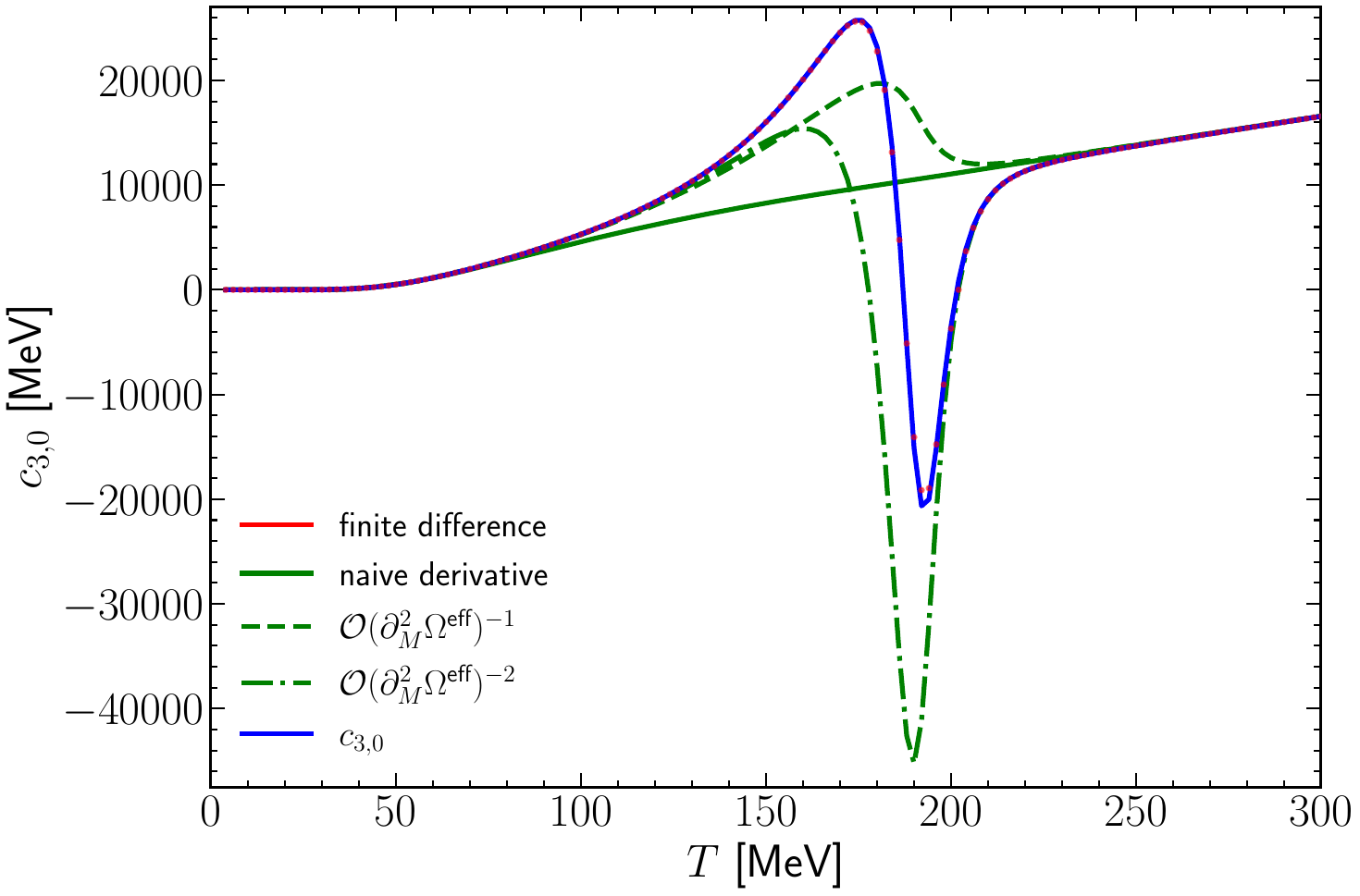}
    \end{minipage}
    \caption{Second- and third-order $T$-derivatives of the pressure at $\mu = 0$:
    the temperature expansion coefficients $c_{2,0} = \partial^2 P/\partial T^2$ (left) and $c_{3,0} = \partial^3 P/\partial T^3$ (right) as functions of $T$. In this case, the numerical calculations (red dots) are more robust and do not show any fluctuations. The symbolic results (blue lines) are in perfect agreement with the numerical ones. Moreover, the naive derivative fails to capture the crossover phase transition behavior, while the symbolic method accurately reproduces it, demonstrating the importance of including constraint-induced corrections.}
    \label{fig:mucoeff}
\end{figure*}

As a final note to this section, we emphasize that our method has been formulated in the context of a grand canonical ensemble; however, it can be straightforwardly adapted to other statistical ensembles. Furthermore, handling cases with more than two internal parameters (e.g., \(T\), \(\mu\), and \(eB\), the external magnetic field) is straightforward: when taking one derivative using the structure of Eq.~\eqref{eq:genstruct}, dependencies on the other parameters naturally drop, as the same happens with \(\mu\) in Eq.~\eqref{eq:jacotder}.
Another important consideration is that certain physical systems may require additional nonlinear constraints on the effective potential. For instance, in neutral quark matter (e.g., see Ref.~\cite{BUBALLA_2005,gholami2024renormalizationgroupconsistenttreatmentcolor}), color and electric charge neutrality are imposed alongside the gap equation. In such scenarios, additional chemical potentials are introduced, with corresponding constraints on their values. These constraints are typically expressed as equations involving derivatives of the effective potential.
The number of constraints determines the number of remaining free (external) parameters, and the symbolic approach presented here can be extended to handle all such constraints. From a mathematical perspective, a gap equation is not different from these additional constraints; thus, the procedure leading to Eq.~\eqref{eq:genstruct} naturally encompasses them, providing a general framework applicable to all such cases.

\section{Applications and Examples}\label{sec:examp}

In this section, we demonstrate the practical implementation and effectiveness of the symbolic approach by applying it to a mean-field NJL model (for a general overview see Ref.~\cite{BUBALLA_2005}). We use the renormalization group (RG) consistent NJL model in  Ref.~\cite{gholami2024renormalizationgroupconsistenttreatmentcolor} for our calculations. As representative quantities, we focus on the speed of sound squared $c_s^2$, the heat capacity at constant volume $c_v$, and higher-order pressure derivatives such as $c_{3,0}$ and $c_{0,3}$.

To compute $c_s^2$ and $c_v$, we require knowledge of several pressure expansion coefficients: $c_{1,0}$, $c_{0,1}$, $c_{2,0}$, $c_{0,2}$, and $c_{1,1}$. Calculation of $c_{3,0}$ and $c_{0,3}$ provide insight into higher-order thermodynamic behavior. We compare these symbolic results with those obtained using traditional numerical differentiation and the naive derivative approach, across various $(T,\mu)$ combinations.

For numerical differentiation, we store the pressure data to arbitrary significant figures and apply a symmetric finite difference scheme. However, this procedure can introduce significant numerical errors, especially for higher-order derivatives where subtractive cancellations occur. Although one might consider fitting splines to the pressure data and then differentiating the fitted function, such spline-based methods can introduce additional artifacts, particularly when higher-order derivatives are required. We therefore recommend avoiding spline fitting for these calculations.

Our initial investigation considers only quark-antiquark pairing in the NJL model. Subsequently, we extend the analysis by including diquark interactions, illustrating how the method works when the model complexity increases.

\begin{figure*}[t]
    \begin{minipage}[t]{0.45\textwidth}		 		
        \centering
        \includegraphics[width=\textwidth]{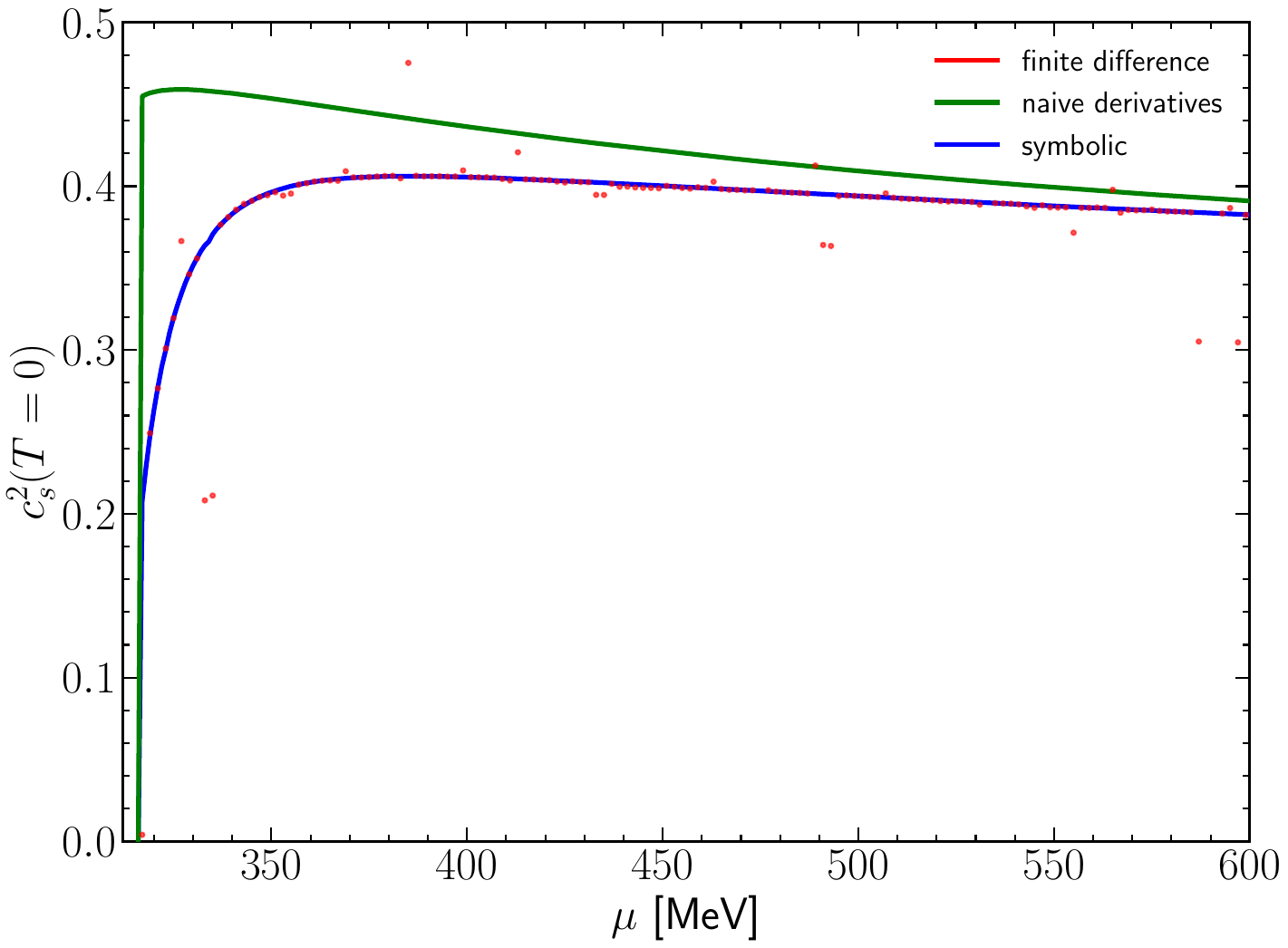}
    \end{minipage}
    \hfill
    \begin{minipage}[t]{0.45\textwidth}
        \centering
        \includegraphics[width=\textwidth]{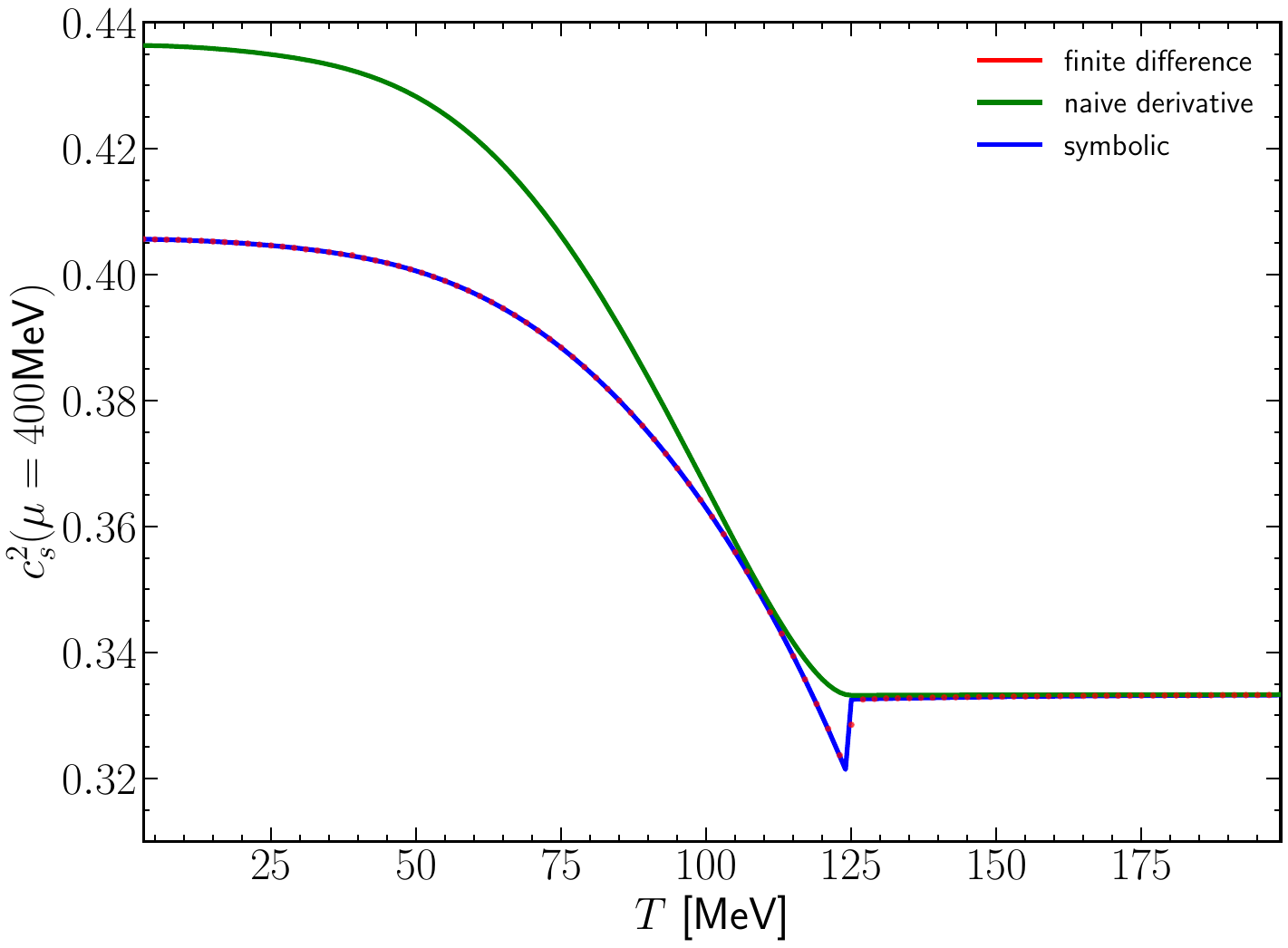}
    \end{minipage}
    \caption{Speed of sound squared $c_s^2$ along two constant slices: $T=0$ (left) and $\mu=400\,\text{MeV}$ (right). In the left panel, the numerical finite difference results (red points) fluctuate significantly due to precision issues and the sensitivity of the method to the chosen step size. By contrast, the symbolic method (blue line) remains stable and accurate. In both panels, using only the naive derivative fails to reproduce the correct behavior. On the right panel, where a second-order phase transition occurs, the naive derivative cannot capture the associated critical features, while the symbolic method does so reliably.}
    \label{fig:csT}	 	
\end{figure*}

\subsection{Two-Flavor NJL Model}

The two-flavor NJL model serves as an ideal testing ground for our Jacobian-based symbolic approach. 
This model is both conceptually simple and well-studied, featuring spontaneous chiral symmetry breaking and known phase transitions, making it a good benchmark for testing our method.

In this model, the effective potential depends on the external parameters $T$ and $\mu$ and involves a single internal parameter, $\chi = M$, the constituent quark mass. Once the physical mass gap, $\bar{M}(T, \mu)$, is determined from the gap equation, the pressure expansion coefficients can be systematically computed. Appendix \ref{app:effective_potential} provides the details of the model, including the explicit expression for the thermodynamic potential and the solutions to its gap equation across different $T$ and $\mu$ combinations.

From Eq.~\eqref{eq:genstruct}, we derive expressions for the first three temperature derivatives $\tilde{c}_{m,0}$ and the first three chemical potential derivatives $\tilde{c}_{0,n}$. These symbolic expansions are presented in Appendix~\ref{app:expansionNJL}. They show that corrections to the naive derivative (see Eq.~\eqref{eq:naivederivative}) appear as negative powers of $\partial_{M}^2 \Omega^{\text{eff}}$ that is, the Jacobian determinants appear in denominators. Using Eq.~\eqref{eq:substi}, the pressure expansion coefficients can be written as
\begin{align}\label{eq:finalcrels}
c_{m,n}(T,\mu) = \frac{\partial^{m+n} P}{\partial T^m \partial \mu^n} = \big[\tilde{c}_{m,n}(T,\mu,M)\big]_{M=\bar{M}}.
\end{align}
As we move to higher derivatives, the expressions grow increasingly complex. In this work, we directly compute the needed derivatives from the NJL effective potential (see Appendix~\ref{app:effective_potential}). For more complicated models or even higher derivatives, an AD implementation can be employed to calculate the symbolic expressions in the $\tilde c_{m,n}$ expansion efficiently.

For numerical comparisons, we use the symmetric finite difference formulas up to third order
\begin{align}
    \frac{\partial P}{\partial x} &\approx \frac{P(x+h) - P(x-h)}{2h}, \label{eq:fd1st}\\
    \frac{\partial^2 P}{\partial x^2} &\approx \frac{P(x+h) - 2P(x) + P(x-h)}{h^2}, \label{eq:fd2nd}\\
    \frac{\partial^3 P}{\partial x^3} &\approx \frac{P(x-2h) - 2P(x-h) + 2P(x+h) - P(x+2h)}{2h^3}. \label{eq:fd3rd}
\end{align}
We store the pressure data to six significant figures. This limited precision introduces numerical errors, most particularly for higher-order derivatives. 
Moreover, the finite difference method is sensitive to the choice of step size $h$, and an inadequate $h$ can amplify numerical noise, causing the computed derivatives to fluctuate. Here, we use the step size of $1$MeV in both temperature and chemical potential direction.

Our symbolic approach avoids these issues by not relying on multiple data points. Instead, we start from Eq.~\eqref{eq:pressure_derivatives_tilde} to obtain the necessary terms. After minimizing the effective potential and determining $\bar{M}(T,\mu)$, we insert the result into Eq.~\eqref{eq:finalcrels}. In Eq.~\eqref{eq:pressure_derivatives_tilde}, the correction terms to the naive derivative appear as inverse powers of $\partial_M^2\Omega^\text{eff}$. We truncate these expansions at different orders to check whether lower-order approximations suffice.

In Fig.~\ref{fig:tempcoeff}, we show $c_{2,0}$ and $c_{3,0}$ at $T=0.1\,\text{MeV}$ for various $\mu$. We choose this low temperature regime because at low temperatures, the Fermi surface approaches a step-like profile, making numerical calculations prone to errors. We compare numerical finite difference results (red dots) with the fully symbolic calculation (blue line). Numerical attempts to extract $c_{0,2}$ suffer from large fluctuations due to limited precision and the sensitivity of the finite difference method to step size $h$, while the symbolic approach remains stable and accurate. As we move away from the phase transition, the full derivative tends towards the naive derivative. Near the phase transition, however, the naive derivative fails to reproduce the correct behavior, and this discrepancy grows at higher orders. For $c_{0,2}$, we find that neglecting the last term in the expansion of Eq.~\eqref{app:c20eq} provides a good approximation.

Figure~\ref{fig:mucoeff} presents $c_{0,2}$ and $c_{0,3}$ at $\mu=0$ as functions of $T$. Here, the numerical method is stable and does not show fluctuations. Still, the naive derivative cannot capture the crossover phase transition, while the symbolic method does so accurately. These corrections to the naive derivative are most crucial in regions of second-order and crossover phase transitions. Unlike the previous case, any approximation seems insufficient to produce the correct result, indicating all terms are necessary for this calculation.

\subsection{Two-Flavor NJL Model With Diquarks}
In the case of diquark pairing added to the previous model, we now have two internal parameters, $\chi_1=M$ and $\chi_2=\Delta$. In this scenario, we are particularly interested in astrophysics relevant quantities such as the speed of sound squared $c_s^2$ and the heat capacity at constant number density $c_v$. For these calculations, only up to second-order pressure coefficients are required (see Eq.~\eqref{eq:cstexp}). Expanding the pressure coefficients to second order using Eqs.~\eqref{eq:firstTder} and \eqref{eq:genstruct} yields the necessary expressions. The complete symbolic terms are listed in Appendix~\ref{app:expansionCSC}.

Using these relations, we can compute $c_s^2$ and $c_v$ at various temperatures and chemical potentials. From Eq.~\eqref{eq:cstexp}, the expression for $c_s^2$ is
\begin{align}\label{eq:csctilde}
    c_s^2(T,\mu) &=\bigg[\bigg( \frac{1}{T \tilde c_{1,0} + \mu \tilde c_{0,1}}\bigg) \\
    &\kern-3em\times\bigg(\frac{2 \tilde c_{1,0} \tilde c_{0,1}\tilde c_{1,1} - \tilde c_{1,0}^2 \tilde c_{0,2} - \tilde c_{0,1}^2 \tilde c_{2,0}}{\tilde c_{1,1}^2 - \tilde c_{2,0}\tilde c_{0,2}}\bigg)\bigg]_{\{M=\bar{M},\Delta=\bar{\Delta}\}},
\end{align}
which, at $T=0$ with $c_{1,0}=c_{1,1}=0$, simplifies to
\begin{align}\label{eq:csctildeT0}
    c_s^2(T=0,\mu) &= \bigg[\frac{\tilde c_{0,1}}{\mu\,\tilde c_{0,2}}\bigg]_{\{T=0;M=\bar{M},\Delta=\bar{\Delta}\}}.
\end{align}
For the heat capacity at constant number density, using Eq.~\eqref{eq;JacobianOmega}, we find
\begin{align}
    c_v(T,\mu)&\equiv T\frac{\partial s}{\partial T}\bigg|_{n}\\
    &= T\bigg[\tilde c_{2,0}-\frac{(\tilde c_{1,1})^2}{\tilde c_{0,2}}\bigg]_{\{M=\bar{M},\Delta=\bar{\Delta}\}},
\end{align}
which goes to zero as $T \to 0$
\begin{align}\label{eq:csvtildeT0}
    c_v(T=0,\mu) &= \lim_{T\to 0}T\big[\tilde c_{2,0}\big]_{\{T=0;M=\bar{M},\Delta=\bar{\Delta}\}}=0.
\end{align}
All these equations require expansion coefficients up to the orders given in Eqs.~\eqref{eq:tildecdiq}.

\begin{figure}[t]
    \centering
    \begin{minipage}[t]{0.45\textwidth}
        \includegraphics[width=\textwidth]{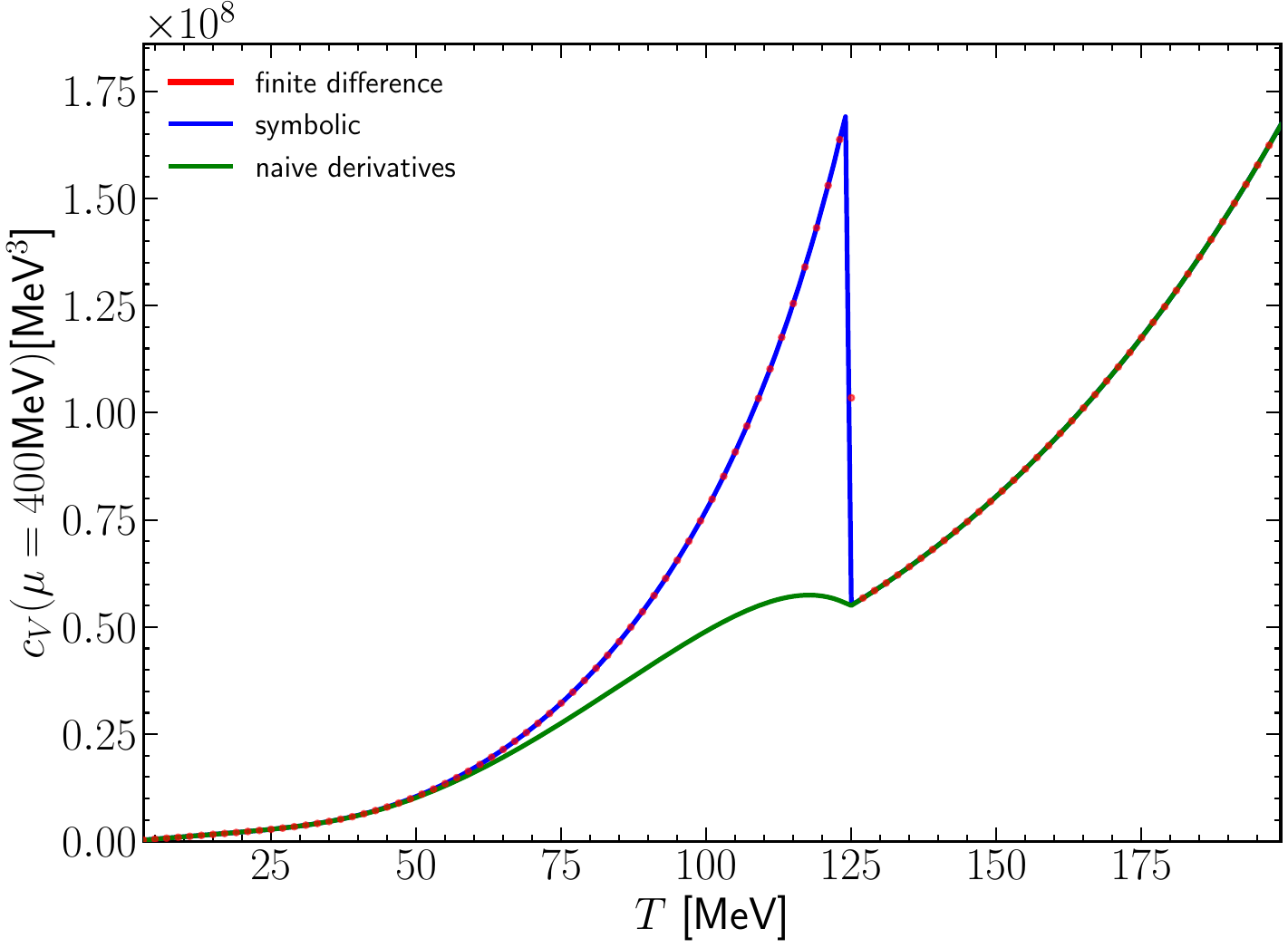}
    \end{minipage}
    \caption{Heat capacity at constant number density $c_v$ along a slice at $\mu=400\,\text{MeV}$. Away from the second-order phase transition region, the naive derivative provides a reasonable approximation. However, as one approaches the critical region, the corrections captured by the symbolic method become essential for accurately determining $c_v$.}
    \label{fig:cVT}	 	
\end{figure}

For the numerical calculations shown in Figs.~\ref{fig:csT} and \ref{fig:cVT}, we increase the numerical precision by storing data with 17 significant digits. The finite difference step size $h$ is set to $1\,\text{MeV}$ in both $T$ and $\mu$ directions. We also examine the speed of sound squared $c_s^2$ using only the naive derivative contributions in $c_{2,0}$, $c_{1,1}$, and $c_{0,2}$, to answer whether this truncation provides acceptable approximations.

Figure~\ref{fig:csT} illustrates the speed of sound squared $c_s^2$ at constant entropy per number density. In the left panel, we display results at $T=0$ for various $\mu$. Even with higher numerical precision, the finite difference method (red points) exhibits fluctuations due to the subtractions in both the numerator and the denominator in Eq.~\eqref{eq:cstexp}. By contrast, the symbolic method (blue line) remains stable and accurate. The naive derivative approximation gives significant deviation from the correct values after the phase transition to the two-flavor color superconductor (2SC) phase. Only at very high chemical potential values they agree.

In the right panel of Fig.~\ref{fig:csT}, we fix $\mu=400\,\text{MeV}$ and vary $T$. Here, the finite difference results are more stable. In this case, the naive derivative approximation also gives significant deviation from the correct values in the 2SC phase. It also fails to capture the second-order phase transition behavior in the vicinity of the critical temperature. The symbolic approach successfully reproduces the expected physical behavior.

In Fig.~\ref{fig:cVT}, we present the heat capacity $c_v$ at constant number density along a slice at $\mu=400\,\text{MeV}$. Away from the second-order phase transition region, the naive derivative provides a reasonable approximation in the low temperatures in the 2SC phase. However, as we approach the critical region, the corrections accounted for by the symbolic method become essential for accurately determining $c_v$.

Overall, these examples demonstrate the advantages of symbolic approach over numerical finite differences. While numerical methods are sensitive to data precision, step size, and spacing, the symbolic method reliably produces smooth and accurate curves, even for the highest-order derivatives considered. Adjusting step sizes and increasing numerical precision may reduce some issues but cannot fully address the instabilities and inaccuracies that emerge near phase transitions, particularly at low temperatures. Moreover, naive derivatives and truncations of expansion coefficients mostly fail to capture the correct phase transition behavior, as they neglect internal parameter dependencies.

\section{Conclusions and Outlook}\label{sec:conclusions}

We have presented a symbolic approach for computing higher-order derivatives of the pressure in mean-field thermal field theories. By deriving symbolic expressions that naturally incorporate the constraints defining the physical internal parameters, our method eliminates the need for multiple-point numerical differentiation and can potentially reduce numerical errors. As a showcase, we calculated pressure expansion coefficients up to third order in the two-flavor NJL model. We also computed the speed of sound and heat capacity in its extensions with diquark pairing and compared these results with finite-difference methods.

This approach is particularly useful for low temperature calculations. In these regimes, numerical differentiation often becomes unstable, but the symbolic method here provides consistent and reliable results. Away from phase transition regions, we find that the naive derivative alone may suffice, simplifying computations. However, near phase transitions or in low-temperature regimes, accounting for the internal parameter dependencies is essential and our symbolic approach gains preference.

In our examples, we have demonstrated that the symbolic method reproduces physically meaningful results even at third order pressure coefficients, while numerical methods struggle. The primary limitation of our current approach is the algebraic complexity that arises when dealing with numerous internal variables or very high-order derivatives. However, this complexity can be mitigated by employing automatic differentiation tools.

It is worth encouraging researchers who calculate models and tabulate thermodynamic quantities to also provide pressure coefficients up to at least the second order. Such data can significantly aid the community by improving the stability and reliability of subsequent calculations.

In the future, we intend to apply our method to more sophisticated models, including the three-flavor NJL model with diquarks of Ref.~\cite{gholami2024renormalizationgroupconsistenttreatmentcolor}. Our results suggest that the symbolic approach can be regarded as a convenient tool for investigating the thermodynamic properties of strongly interacting matter within the mean-field approximation, delivering stable results in challenging regimes.

\onecolumngrid
\section*{Acknowledgement}
I extend my gratitude to Michael Buballa, Marco Hofmann, Lennart Kurth, Ugo Mire, 
D\'ebora Mroczek, Jaki Noronha-Hostler, and Jonas Turnwald for their valuable feedback on this 
manuscript.
 I am especially thankful to Ugo Mire and Andreas Cardona for implementing this method in their calculations and providing further insights. I also acknowledge Marco Hofmann and Lennart Kurth for their helpful suggestions on improving the notation used in this text. I thank Dirk Rischke for recognizing the potential of this method and encouraging me to pursue its development. I acknowledge support from the Deutsche Forschungsgemeinschaft (DFG, German Research Foundation) 
through the CRC-TR211 'Strong-interaction matter under extreme conditions' project number 315477589 – TRR 211. 

\section*{Supplementary Material}

Two Mathematica notebooks accompanying this manuscript are available in the repository of Ref.~\cite{Gholami_Mean-field_symblolic_pressure_2025}. The first notebook provides a script for deriving symbolic expressions for arbitrary \((n+m)\)-th derivatives of the pressure by an automatized implementation of Eq.~\eqref{eq:genstruct}. The second notebook demonstrates the application of this approach to the two-flavor NJL model, with calculations consistent with those presented in Sec.~\ref{sec:examp} of this manuscript.

\appendix
\section{alternative approach}\label{app:alternative}
Having only one internal parameter for the effective potential
\begin{eqnarray}
    \Omega^\text{eff}\left(T,\mu,\chi\right),
\end{eqnarray}
the physical solutions $\bar \chi$ are determined by the condition 
\begin{eqnarray}
    \frac{d}{d \chi}\Omega^\text{eff}\left(T,\mu,\chi\right)\big|_{\chi=\bar \chi}=0.
\end{eqnarray}
Inserting the physical solutions at all $T$ and $\mu$ values back to the effective potential give the thermodynamic potential
\begin{eqnarray}
    \Omega(T,\mu)=\Omega^\text{eff}\left(T,\mu,\bar \chi(T,\mu)\right).
\end{eqnarray}
Suppose we are interested in the second derivative of the thermodynamic potential with respect to $T$. The first derivative with respect to $T$ gives
\begin{eqnarray}\label{eq:appfirstchain}
    \frac{d}{d T} \Omega(T,\mu)=
    \frac{d}{d T}\Omega^\text{eff}(T,\mu,\bar \chi(T,\mu))=
    \left[\frac{\partial}{\partial T}\Omega^\text{eff}(T,\mu, \chi)\right]_{\chi=\bar\chi}+\frac{\partial \bar \chi}{\partial T}\bigg[\frac{\partial \Omega^\text{eff}(T,\mu, \chi)}{\partial  \chi}\bigg]_{\chi=\bar\chi}.
\end{eqnarray}
The last term vanishes for the physical solutions $\bar{\chi}$, as defined in Eq.~\eqref{eq:condition}.
For the second derivative we get
\begin{eqnarray}{\label{app:expansion}}
    \frac{\partial^2}{\partial T^2} \Omega(T,\mu)&=&\nonumber\frac{\partial}{\partial T} \left(\frac{\partial}{\partial T}\Omega^\text{eff}(T,\mu,\bar \chi(T,\mu))\right)\\
    &=&\frac{\partial ^2}{\partial T^2}\Omega^\text{eff}(T,\mu, \bar \chi)+\frac{\partial \bar \chi}{\partial T}\bigg[\frac{\partial^2 \Omega^\text{eff}(T,\mu, \chi)}{\partial \chi \partial T}\bigg]_{\chi=\bar\chi}.
\end{eqnarray}
One can get a symbolic expression for derivative of the internal parameter with respect to the external parameters. For this, we can use the fact that for all $T$ and $\mu$, the condition $\partial_\chi\Omega\big|_{\chi=\bar\chi}=0$ holds. We have
\begin{eqnarray}
  \frac{\partial}{\partial T} \bigg( \bigg[\frac{\partial\Omega^{\text{eff}}}{\partial  \chi}\bigg]_{\chi=\bar\chi}\bigg)\bigg|_\mu&=&0,\\
    \frac{\partial}{\partial \mu} \bigg(\bigg[\frac{\partial\Omega^{\text{eff}}}{\partial  \chi}\bigg]_{\chi=\bar\chi}\bigg)\bigg|_T&=&0.\label{app:altn}
\end{eqnarray}
The first line  expands as
\begin{eqnarray}
    \bigg[\frac{\partial^2\Omega_{\text{eff}}}{\partial T\partial  \chi}+\frac{\partial^2\Omega_{\text{eff}}}{\partial \chi^2}\frac{\partial \chi}{\partial T}\bigg]_{\chi=\bar\chi}=0.
\end{eqnarray}
The above expression gives us the derivative needed in Eq.~\eqref{app:expansion} and translates it to derivatives of $\Omega_{\text{eff}}$. We have
\begin{eqnarray}
    \frac{\partial\bar \chi}{\partial T}=-\bigg[\left(\frac{\partial^2\Omega_{\text{eff}}}{\partial \chi^2}\right)^{-1}\frac{\partial^2\Omega_{\text{eff}}}{\partial T\partial \chi}\bigg]_{\chi=\bar\chi}.
\end{eqnarray}
Plugging this back to Eq.~\eqref{app:expansion}, we get
\begin{eqnarray}
    c_{2,0}=-\frac{\partial^2}{\partial T^2} \Omega(T,\mu)=-\bigg[\Bigg(
\frac{\partial^2 \Omega^\text{eff}(T,\mu, \chi)}{\partial T \partial \mu} - \frac{ \dfrac{\partial^2 \Omega^\text{eff}(T,\mu, \chi)}{\partial \mu \partial \chi} \cdot \dfrac{\partial^2 \Omega^\text{eff}(T,\mu, \chi)}{\partial T \partial \chi} }{ \dfrac{\partial^2 \Omega^\text{eff}(T,\mu, \chi)}{\partial \chi^2} }
\Bigg) \bigg]_{\chi = \bar{\chi}},
\end{eqnarray}
which is the same expression that we get in Eq.~\eqref{eq:stjac} using our proposed method. One can easily check that the expansion in Eq.~\eqref{app:altn} can be used to get the expression for $\frac{\partial^2\Omega}{\partial \mu^2}$. In the general case, with having multiple internal parameters $\vect\chi$, one can solve a system of equations
\begin{eqnarray}
   \bigg\{ \frac{\partial}{\partial T}\bigg(\bigg[\frac{\partial\Omega^{\text{eff}}}{\partial  \chi_i}\bigg]_{\vect\chi=\bar{\vect\chi}}\bigg)\bigg|_\mu&=&0\bigg\}_{\text{for all $i$}},\\
    \bigg\{ \frac{\partial}{\partial \mu}\bigg(\bigg[\frac{\partial\Omega^{\text{eff}}}{\partial  \chi_i}\bigg]_{\vect\chi=\bar{\vect\chi}}\bigg)\bigg|_T&=&0\bigg\}_{\text{for all $i$}},
\end{eqnarray}
to get all the expressions necessary for calculating the derivatives symbolically. However, using the Jacobians in our proposed method automates this process, making it more convenient to use.

\section{model details}\label{app:effective_potential}

In this appendix, we show the model details of our calculations in Sec.~\ref{sec:examp}. We use an RG consistent NJL model, introduced in Ref.~\cite{gholami2024renormalizationgroupconsistenttreatmentcolor}.

\subsection{Two-Flavor NJL Model}\label{app:NJLmodel}

The mean-field effective potential for the two-flavor Nambu--Jona-Lasinio (NJL) model is expressed as

\begin{align}\label{eq:effective_potential_NJL}
\Omega^{\text{eff}}(T, \mu, M) &= - \int_{0}^{\infty} \frac{6 p^2}{\pi^2} \left[ T \ln \left(1 + e^{-\frac{\sqrt{M^2 + p^2} - \mu}{T}} \right) + T \ln \left(1 + e^{-\frac{\sqrt{M^2 + p^2} + \mu}{T}} \right) \right] \, dp \nonumber \\
&\quad - \int_{0}^{\Lambda} \frac{6 p^2 \sqrt{M^2 + p^2}}{\pi^2} \, dp + \frac{(M - m_0)^2}{4 G},
\end{align}
where the first term represents the medium contribution. For RG consistency, we extend this integration to infinity. The second term corresponds to the vacuum contribution, regulated by a momentum cutoff scale $\Lambda$. The last term is the potential part of the effective action. Here, $m_0$ is the bare quark mass, and $G$ is the NJL coupling constant. The values of these parameters are listed in Table~\ref{tab:parameters}. The quark mass $M$ is related to the order parameter for the chiral phase transition and is determined by solving the gap equation $\partial_M\Omega^\text{eff}=0$.

\begin{figure*}[t]
    \begin{minipage}[t]{0.45\textwidth}		 		
        \includegraphics[width=\textwidth]{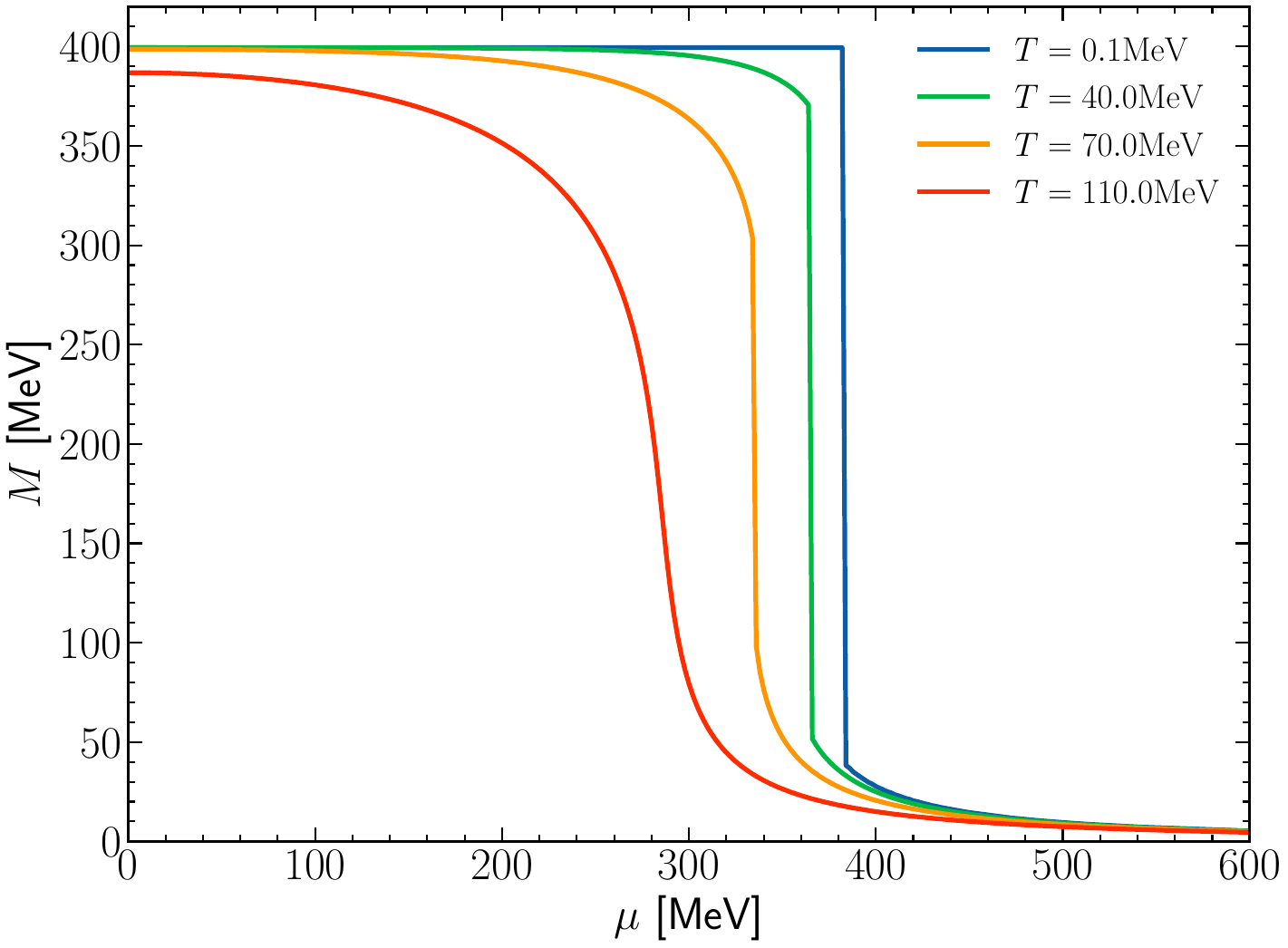}
    \end{minipage}
    \hfill
    \begin{minipage}[t]{0.45\textwidth}
        \includegraphics[width=\textwidth]{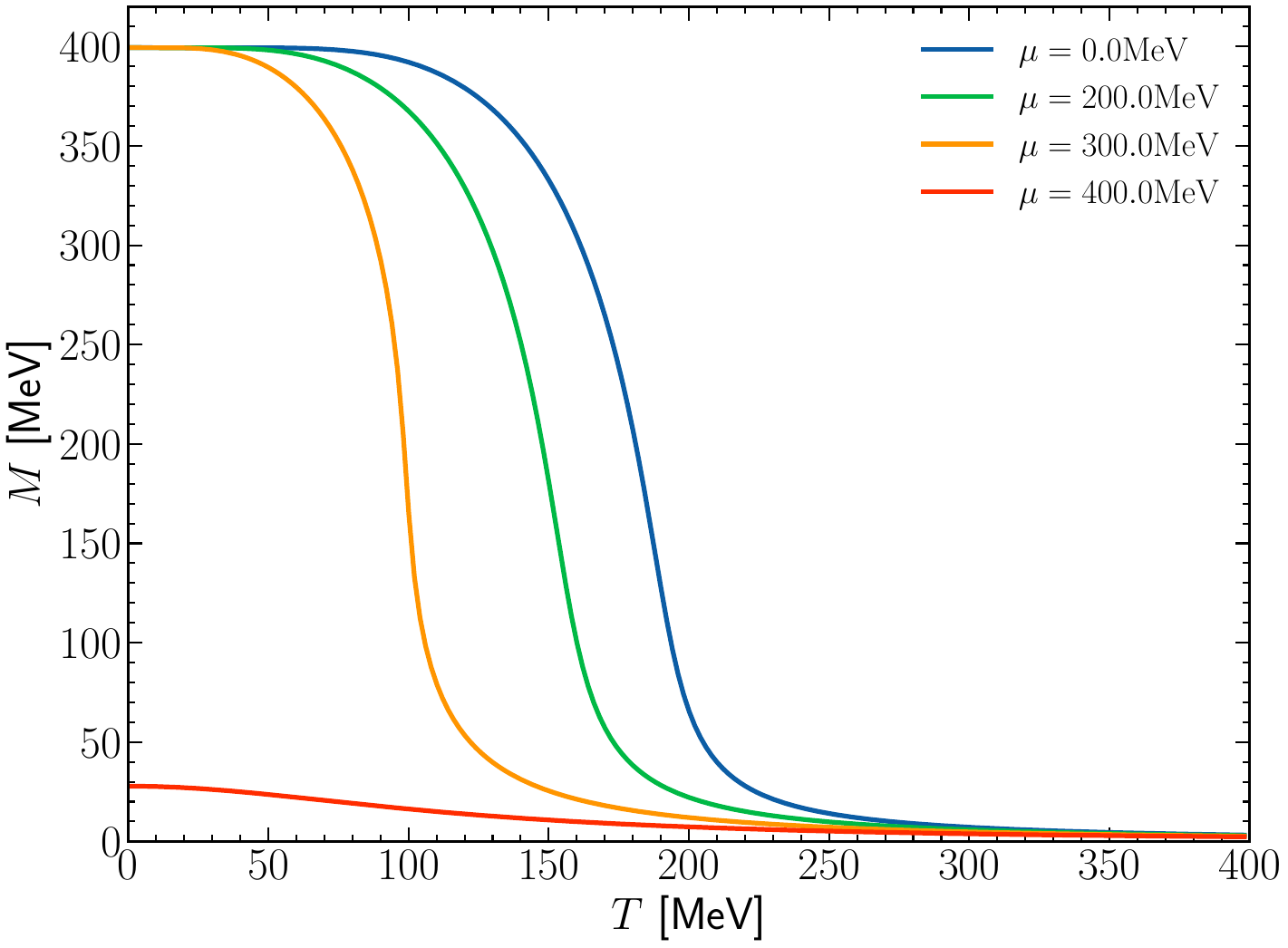}
    \end{minipage}
    \caption{Quark mass \( M \) as functions of chemical potential \( \mu \) (left) and temperature \( T \) (right). At low temperatures, increasing \( \mu \) induces a sharp phase transition to an approximately chirally restored phase. At high temperatures, this phase transition becomes second order in both \( \mu \) and \( T \) directions.}
    \label{fig:Mvs}	 	
\end{figure*}

\begin{table}[h]
    \centering
    \label{tab:parameters}
    \begin{tabular}{|c|c|c||c|}
        \hline
        $\Lambda$ & $G$ & $m_0$&$H$ \\ \hline
        587.9 MeV & $2.44/\Lambda^2$ MeV$^{-2}$ & 5.6 MeV&$2.44/\Lambda^2$ MeV$^{-2}$ \\ \hline
    \end{tabular}
        \caption{Parameter values for the two-flavor NJL model. Values taken from Ref.~\cite{BUBALLA_2005}.}
\end{table}
Figure \ref{fig:Mvs} illustrates the quark mass \( M \) for different combinations of temperature \( T \) and chemical potential \( \mu \). At low temperatures, increasing the chemical potential leads to a sharp phase transition to an approximately chirally restored phase. Conversely, at high temperatures, this phase transition becomes second order in both \( \mu \) and \( T \) directions.
\subsection{NJL Model with Diquarks}

Incorporating diquarks into the system, the dispersion relations for paired quark combinations in color-flavor anti-triplet channel quasi-particles and quasi-anti-particles are given by

\begin{eqnarray}
\omega_{\mp} = \sqrt{(\sqrt{p^2 + M^2} \mp \mu)^2 + \Delta^2}.
\end{eqnarray}
Here, $M$ is the quark mass and $\Delta$ is the diquark condensate. 
The mean-field effective potential for this model, including diquark pairing, is expressed as

\begin{align}\label{eq:effective_potential_diquarks}
\Omega^{\text{eff}}(T, \mu, M, \Delta) &= -\frac{2}{\pi^2} \int_{0}^{\infty} p^2 \bigg[ 
    T \ln \left(1 + e^{-\frac{\sqrt{M^2 + p^2} - \mu}{T}} \right) 
    + T \ln \left(1 + e^{-\frac{\sqrt{M^2 + p^2} + \mu}{T}} \right) \nonumber \\
&\quad + 2 T \ln \left(1 + e^{-\frac{\omega_-}{T}} \right) 
    + 2 T \ln \left(1 + e^{-\frac{\omega_+}{T}} \right) 
    + \omega_- + \omega_+  
    - 2\sqrt{p^2 + M^2 + \Delta^2} 
\bigg] dp \\
&\quad -\frac{2}{\pi^2} \int_{0}^{\Lambda} p^2 \bigg[ 
    \sqrt{M^2 + p^2} 
    + 2\sqrt{p^2 + M^2 + \Delta^2} 
\bigg] dp \nonumber \\
&\quad + \frac{2}{\pi^2} \mu^2 \Delta^2 \int_{\Lambda}^{\infty} p^2 \left[ 
    \frac{1}{\left(p^2 + \Delta^2 + M^2\right)^{3/2}} 
\right] dp \nonumber \\
&\quad + \frac{(M - m_0)^2}{4 G} + \frac{\Delta^2}{4 H},
\end{align}

where the first two lines represent the medium contributions, integrated to infinity to ensure renormalization group consistency. The third term is the vacuum contribution, regulated by the momentum cutoff $\Lambda$. The fourth line accounts for medium renormalization contributions in the massive scheme (for details of this scheme, see Ref.~\cite{gholami2024renormalizationgroupconsistenttreatmentcolor}). The final line is the potential part of the effective action. 
We use the same values for $\Lambda$, $m_0$ and $G$ from table \ref{tab:parameters}. In addition, we have $H$ being the diquark coupling strength. The value for $H$ is listed in table~\ref{tab:parameters}.
The quark mass $M$ and diquark condensate $\Delta$ serve as order parameters for the chiral and color superconductivity phase transitions, respectively. They are determined by simultaneously solving the gap equations $\partial_M \Omega^{\text{eff}} = 0$ and $\partial_\Delta \Omega^{\text{eff}} = 0$.

\begin{figure*}[t]
    \begin{minipage}[t]{0.45\textwidth}		 		
        \includegraphics[width=\textwidth]{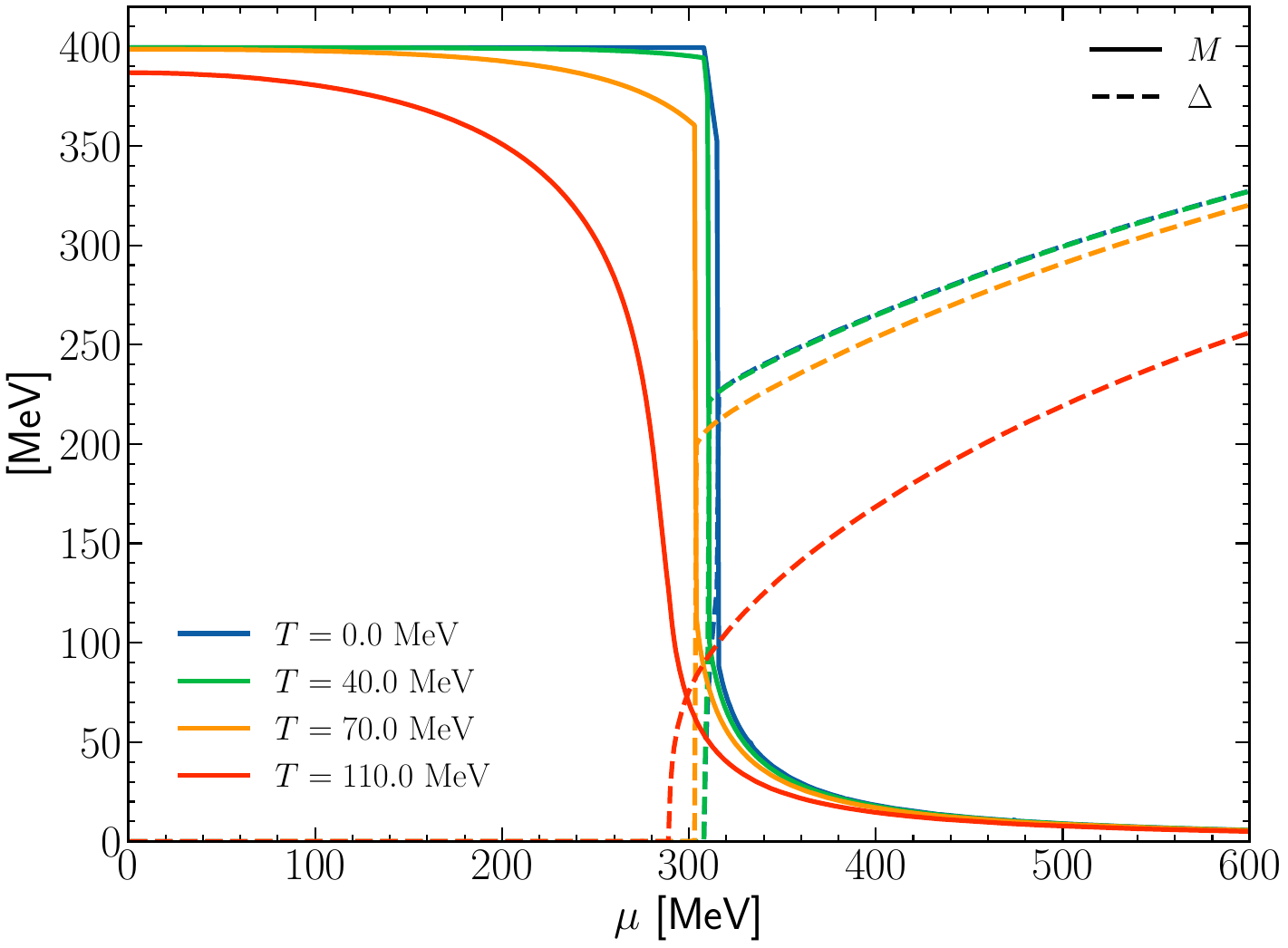}
    \end{minipage}
    \hfill
    \begin{minipage}[t]{0.45\textwidth}
        \includegraphics[width=\textwidth]{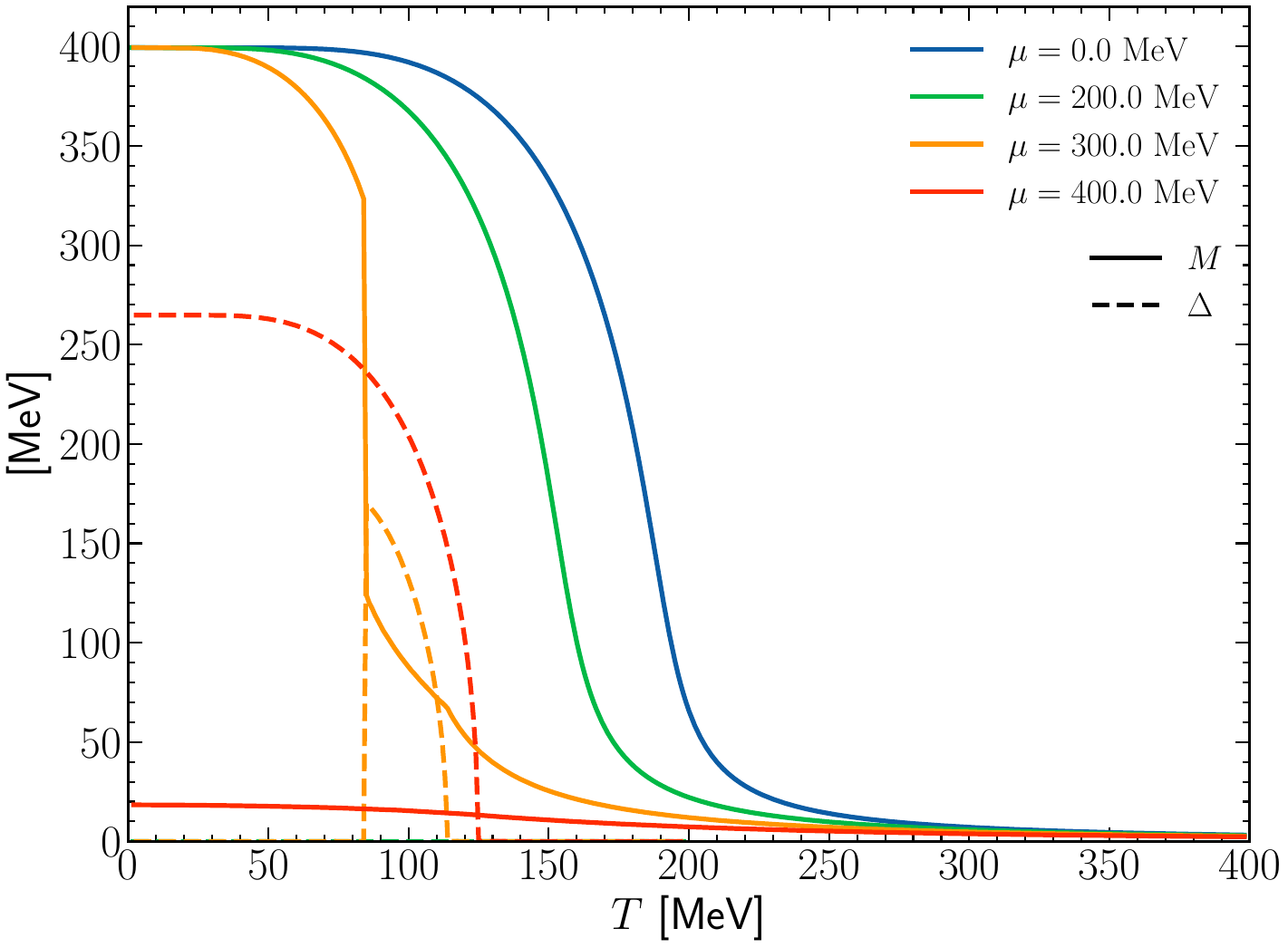}
    \end{minipage}
    \caption{Quark mass \( M \) as functions of chemical potential \( \mu \) (left) and temperature \( T \) (right). At low temperatures, increasing \( \mu \) induces a sharp phase transition to an approximately chirally restored phase. At high temperatures, this phase transition becomes second order in both \( \mu \) and \( T \) directions.}
    \label{fig:dvs}	 	
\end{figure*}

Figure \ref{fig:dvs} illustrates the behavior of the diquark condensate $\Delta$ (dashed line) and the quark mass $M$ (solid line) for different combinations of temperature $T$ and chemical potential $\mu$. At low temperatures, increasing the chemical potential leads to a sharp first-order phase transition to a two-flavor superconducting (2SC) phase, characterized by a sudden drop in $M$ and a nonzero value for $\Delta$. At high temperatures, this phase transition becomes second-order, transitioning to the chirally restored phase without diquarks. Additionally, at high chemical potentials, with going to high temperatures, the diquark condensate melts away, resulting in a second-order phase transition to the chirally restored phase.


\section{\(\tilde c_{m,n}\) expansions}
\subsection*{NJL model}\label{app:expansionNJL}
In the RG consistent NJL model, the pressure expansion coefficients $\tilde{c}_{m,n}$ are derived from the effective potential $\Omega^{\text{eff}}$, defined in Eq.~\eqref{eq:effective_potential_NJL}. The $\tilde c$ coefficients up to third order for both temperature and chemical potential derivatives are given by
\begin{subequations}\label{eq:pressure_derivatives_tilde}
    \begin{align}
    \tilde{c}_{1,0}(T,\mu, M) &= -\partial_T \Omega^\text{eff}, \\
    \tilde{c}_{2,0}(T,\mu, M) &= -\partial_T^2 \Omega^\text{eff} 
        + \frac{\left( \partial_T \partial_M \Omega^\text{eff} \right)^2}{\partial_M^2 \Omega^\text{eff}}, \\
        \tilde{c}_{3,0}(T,\mu, M) &=-\partial_T^3 \Omega^{\text{eff}} + \frac{ 3 \partial_T \partial_{M} \Omega^{\text{eff}} \, \partial_T^2 \partial_{M} \Omega^{\text{eff}} }{ \partial_{M}^2 \Omega^{\text{eff}} }
- \frac{ 3 \left( \partial_T \partial_{M} \Omega^{\text{eff}} \right)^2 \partial_T \partial_{M}^2 \Omega^{\text{eff}} }{ \left( \partial_{M}^2 \Omega^{\text{eff}} \right)^2 }
+ \frac{ \partial_{M}^3 \Omega^{\text{eff}} \left( \partial_T \partial_{M} \Omega^{\text{eff}} \right)^3 }{ \left( \partial_{M}^2 \Omega^{\text{eff}} \right)^3 } ,\\\nonumber\\
        \tilde{c}_{0,1}(T,\mu, M) &= -\partial_\mu \Omega^\text{eff}, \\ 
    \tilde{c}_{0,2}(T,\mu, M) &=- \partial_\mu^2 \Omega^\text{eff} 
        + \frac{ \left( \partial_\mu \partial_M \Omega^\text{eff} \right)^2 }{ \partial_M^2 \Omega^\text{eff} },\\   
 \tilde{c}_{0,3}(T,\mu, M) &=-\partial_\mu^3 \Omega^{\text{eff}} + \frac{ 3 \partial_\mu \partial_{M} \Omega^{\text{eff}} \, \partial_\mu^2 \partial_{M} \Omega^{\text{eff}} }{ \partial_{M}^2 \Omega^{\text{eff}} }
- \frac{ 3 \left( \partial_\mu \partial_{M} \Omega^{\text{eff}} \right)^2 \partial_\mu \partial_{M}^2 \Omega^{\text{eff}} }{ \left( \partial_{M}^2 \Omega^{\text{eff}} \right)^2 }
+ \frac{ \partial_{M}^3 \Omega^{\text{eff}} \left( \partial_\mu \partial_{M} \Omega^{\text{eff}} \right)^3 }{ \left( \partial_{M}^2 \Omega^{\text{eff}} \right)^3 } .
\end{align}
\end{subequations}
Here we only show the pure \(T\) or \(\mu\) derivatives used in Sec.~\ref{sec:examp}, but not the mixed ones. The second mixed derivative and third mixed derivatives are
\begin{align}
        \tilde{c}_{1,1}(T,\mu, M) &= -\partial_T \partial_\mu \Omega^\text{eff} 
        + \frac{ \left( \partial_T \partial_M \Omega^\text{eff} \right) \left( \partial_\mu \partial_M \Omega^\text{eff} \right) }{ \partial_M^2 \Omega^\text{eff} }\\
    c_{2,1}(T, \mu, M) &= - \partial_T^2 \partial_\mu \Omega^\text{eff} + \frac{2 \, \partial_T \partial_\mu \partial_\chi \Omega^\text{eff} \, \partial_T \partial_\chi \Omega^\text{eff} + \partial_\mu \partial_\chi \Omega^\text{eff} \, \partial_T^2 \partial_\chi \Omega^\text{eff}}{\partial_\chi^2 \Omega^\text{eff}}\nonumber \\
    &\quad - \frac{\partial_T \partial_\chi \Omega^\text{eff} \big(2 \partial_T \partial_\chi^2 \Omega^\text{eff} \, \partial_\mu \partial_\chi \Omega^\text{eff} + \partial_T \partial_\chi \Omega^\text{eff} \, \partial_\mu \partial_\chi^2 \Omega^\text{eff} \big)}{(\partial_\chi^2 \Omega^\text{eff})^2} +\frac{\partial_\chi^3 \Omega^\text{eff} \, (\partial_T \partial_\chi \Omega^\text{eff})^2 \, \partial_\mu \partial_\chi \Omega^\text{eff}}{(\partial_\chi^2 \Omega^\text{eff})^3}\\
    c_{1,2}(T, \mu, M) &= -\partial_T \partial_\mu^2 \Omega^\text{eff} + \frac{2 \, \partial_T \partial_\mu \partial_\chi \Omega^\text{eff} \, \partial_\mu \partial_\chi \Omega^\text{eff} + \partial_T \partial_\chi \Omega^\text{eff} \, \partial_\mu^2 \partial_\chi \Omega^\text{eff}}{\partial_\chi^2 \Omega^\text{eff}} \nonumber\\
    &\quad   - \frac{\partial_\mu \partial_\chi \Omega^\text{eff} \big(\partial_T \partial_\chi^2 \Omega^\text{eff} \, \partial_\mu \partial_\chi \Omega^\text{eff} + 2 \partial_T \partial_\chi \Omega^\text{eff} \, \partial_\mu \partial_\chi^2 \Omega^\text{eff}\big)}{(\partial_\chi^2 \Omega^\text{eff})^2} + \frac{\partial_\chi^3 \Omega^\text{eff} \, \partial_T \partial_\chi \Omega^\text{eff} \, (\partial_\mu \partial_\chi \Omega^\text{eff})^2}{(\partial_\chi^2 \Omega^\text{eff})^3}
\end{align}
The expressions required for the above calculations are as follows
\begin{description}
    \item [ First derivatives (3 terms)]
    \(\partial_T \Omega^\text{eff}, \partial_\mu \Omega^\text{eff}, \partial_M \Omega^\text{eff}\).
    
    \item[ Second derivatives (6 terms)]
    \(\partial_T^2 \Omega^\text{eff}, \partial_\mu^2 \Omega^\text{eff}, \partial_M^2 \Omega^\text{eff}, \partial_T \partial_\mu \Omega^\text{eff}, \partial_T \partial_M \Omega^\text{eff}, \partial_\mu \partial_M \Omega^\text{eff}\).
    
    \item [ Third derivatives (10 terms)]
    \begin{align*}
    &\partial_T^3 \Omega^\text{eff},  \partial_\mu^3 \Omega^\text{eff},  \partial_M^3 \Omega^\text{eff},  \partial_T^2 \partial_\mu \Omega^\text{eff},  \partial_T^2 \partial_M \Omega^\text{eff}, \partial_\mu^2 \partial_T \Omega^\text{eff},  \partial_\mu^2 \partial_M \Omega^\text{eff},  \partial_M^2 \partial_T \Omega^\text{eff},  \partial_M^2 \partial_\mu \Omega^\text{eff},  \partial_T \partial_\mu \partial_M \Omega^\text{eff}.
    \end{align*}
\end{description}
The total number of unique derivatives required up to the third order derivatives is therefore 19.
\subsection*{NJL model with diquarks}\label{app:expansionCSC}
When diquark pairing is included in the NJL model, an additional internal parameter $\Delta$—the diquark condensate—enters the effective potential. $\tilde c$ coefficients up to second order are presented below

\begin{subequations}\label{eq:tildecdiq}
    \begin{align}
\tilde{c}_{1,0}(T,\mu,M,\Delta) &= -\partial_T \Omega^{\text{eff}}, \\[8pt]
\tilde{c}_{0,1}(T,\mu,M,\Delta) &= -\partial_\mu \Omega^{\text{eff}}, \\[8pt]
\tilde{c}_{1,1}(T,\mu,M,\Delta) 
&= - (\partial_T \partial_\mu \Omega^{\text{eff}})-\frac{1}{
(\partial_M \partial_\Delta \Omega^{\text{eff}})^2
- (\partial_\Delta^2 \Omega^{\text{eff}})(\partial_M^2 \Omega^{\text{eff}})
}
\Big[
 \nonumber\\[6pt]
&\qquad ((\partial_M^2 \Omega^{\text{eff}})(\partial_\mu \partial_\Delta \Omega^{\text{eff}})
- (\partial_M \partial_\Delta \Omega^{\text{eff}})(\partial_\mu \partial_M \Omega^{\text{eff}}))
(\partial_T \partial_\Delta \Omega^{\text{eff}})\nonumber\\[6pt]
&\quad+ \bigl(
 (\partial_\Delta^2 \Omega^{\text{eff}})(\partial_\mu \partial_M \Omega^{\text{eff}})-(\partial_M \partial_\Delta \Omega^{\text{eff}})(\partial_\mu \partial_\Delta \Omega^{\text{eff}})\bigr)
(\partial_T \partial_M \Omega^{\text{eff}})
\Big]
\;, 
 \\[8pt]
\tilde{c}_{2,0}(T,\mu,M,\Delta) &=-\partial_T^2 \Omega^{\text{eff}}-
\frac{
(\partial_M^2 \Omega^{\text{eff}})(\partial_T \partial_\Delta \Omega^{\text{eff}})^2
- 2 (\partial_M \partial_\Delta \Omega^{\text{eff}})
(\partial_T \partial_\Delta \Omega^{\text{eff}})
(\partial_T \partial_M \Omega^{\text{eff}})
+ (\partial_\Delta^2 \Omega^{\text{eff}})
(\partial_T \partial_M \Omega^{\text{eff}})^2
}{
(\partial_M \partial_\Delta \Omega^{\text{eff}})^2
- (\partial_\Delta^2 \Omega^{\text{eff}})(\partial_M^2 \Omega^{\text{eff}})
}
, \\[8pt]
\tilde{c}_{0,2}(T,\mu,M,\Delta) &=- \partial_\mu^2 \Omega^{\text{eff}}-
\frac{
(\partial_M^2 \Omega^{\text{eff}})(\partial_\mu \partial_\Delta \Omega^{\text{eff}})^2
- 2 (\partial_M \partial_\Delta \Omega^{\text{eff}})
(\partial_\mu \partial_\Delta \Omega^{\text{eff}})
(\partial_\mu \partial_M \Omega^{\text{eff}})
+ (\partial_\Delta^2 \Omega^{\text{eff}})
(\partial_\mu \partial_M \Omega^{\text{eff}})^2
}{
(\partial_M \partial_\Delta \Omega^{\text{eff}})^2
- (\partial_\Delta^2 \Omega^{\text{eff}})(\partial_M^2 \Omega^{\text{eff}})
}\label{app:c20eq}
.
\end{align}
\end{subequations}
The expressions required for these calculations are as follows
\begin{description}
    \item[ First derivatives (4 terms)]  \(\partial_T \Omega^\text{eff}, \partial_\mu \Omega^\text{eff}, \partial_M \Omega^\text{eff}, \partial_\Delta \Omega^\text{eff}\).
    \item[ Second derivatives (10 terms)] 
    \begin{align*}
    \partial_T^2 \Omega^\text{eff}, \partial_\mu^2 \Omega^\text{eff}, \partial_M^2 \Omega^\text{eff}, \partial_\Delta^2 \Omega^\text{eff}, \partial_T \partial_\mu \Omega^\text{eff}, \partial_T \partial_M \Omega^\text{eff}, \partial_T \partial_\Delta \Omega^\text{eff}, \partial_\mu \partial_M \Omega^\text{eff}, \partial_\mu \partial_\Delta \Omega^\text{eff}, \partial_M \partial_\Delta \Omega^\text{eff}.
    \end{align*}
\end{description}

Thus, the total number of unique derivatives required up to the second order derivatives is 14.

\twocolumngrid

\bibliographystyle{apsrev4-1}
\bibliography{references.bib}

\end{document}